\documentclass[aps,twocolumn,pra,superscriptaddress]{revtex4-2}

\usepackage{graphicx}  
\usepackage{subfigure}
\usepackage{multirow}
\usepackage{xeCJK}
\usepackage{fancyhdr}
\usepackage{longtable}
\usepackage{parskip}
\usepackage[T1]{fontenc}
\usepackage{placeins}
\usepackage{bm}        
\usepackage{amsfonts}  
\usepackage{amsmath}   
\usepackage{amssymb}   
\usepackage{url}
\usepackage{lipsum}
\usepackage{hyperref}
\hypersetup{
    colorlinks=true,
    linkcolor=blue,
    filecolor=blue,
    urlcolor=blue,
    citecolor=blue,
}

\newcommand{\beq}{\begin{equation}}
	\newcommand{\eeq}{\end{equation}}
\newcommand{\beqa}{\begin{eqnarray}}
	\newcommand{\eeqa}{\end{eqnarray}}

\newcommand{\pwisein}{\left\{ \begin{array}{ll}}
	\newcommand{\pwiseout}{\end{array}\right.}
\newcommand{\ket}[1]{\left| #1 \right\rangle}

\newcommand{\melb}[3]{\big\langle #1\big|#2 \big| #3\big \rangle}

\newcommand{\dx}{\mathrm{d}}
\begin{document}

\title{Hyperfine-resolved laser excitation and detection of nuclear isomer in trapped $^{229}$Th$^{3+}$ ions}

\author{Wu Wang (王武)}
\email[Corresponding author: ]{wangwu@itp.ac.cn}
\affiliation {\it Institute of Theoretical Physics, Chinese Academy of Sciences, Beijing 100190, China}

\author{Ke Zhang (张科)}
\email[Corresponding author: ]{kezhangvv@bit.edu.cn}
\affiliation{\it Center for Photonic Quantum Precision Measurement, Advanced Research Institute of Multidisciplinary Science, Beijing Institute of Technology, Beijing 100081, China}
\affiliation{\it Beijing Institute of Technology, State Key Laboratory of Environment Characteristics and Effects for Near-space,
	Beijing 100081, China}
\affiliation{\it Johannes
	Gutenberg-University Mainz, 55128 Mainz, Germany}
	
\author{Ke-Mi Xu (徐可米)}
\affiliation{\it Center for Photonic Quantum Precision Measurement, Advanced Research Institute of Multidisciplinary Science, Beijing Institute of Technology, Beijing 100081, China}
\affiliation{\it Beijing Institute of Technology, State Key Laboratory of Environment Characteristics and Effects for Near-space,
	Beijing 100081, China}

\author{Jin-Bo Hu (胡锦波)}
\affiliation {\it Beijing Computational Science Research Center, Beijing 100193, China}
	
\author{Shan-Gui Zhou (周善贵)}
\email[Corresponding author: ]{sgzhou@itp.ac.cn }
\affiliation {\it Institute of Theoretical Physics, Chinese Academy of Sciences, Beijing 100190, China}
\affiliation {\it School of Physical Sciences, University of Chinese Academy of Sciences, Beijing 100049, China}
\affiliation {\it School of Nuclear Science and Technology, University of Chinese Academy of Sciences, Beijing 100049, China}

\date{\today}

\begin{abstract}
		We present a comprehensive theoretical investigation of
	hyperfine-resolved excitation and detection of the low-energy isomeric state of $^{229}$Th in trapped $^{229}\mathrm{Th}^{3+}$ ions. Using a quantum master equation approach, we quantitatively analyze the dependence of the isomeric population on laser linewidth, detuning, and irradiation time, showing that their proper matching is essential for efficient excitation. Going beyond earlier conceptual discussions of electronic-fluorescence-based nuclear-state detection, we propose two concrete nuclear-state detection schemes based on three hyperfine-resolved electronic fluorescence channels at 690, 984, and 1088 nm. Our quantitative analysis shows that, for 50 ions, the 690- and 984-nm scheme yields detectable photon count rates on the order of $10^3~\mathrm{s}^{-1}$  at 690 nm and  $10^4~\mathrm{s}^{-1}$ at 984 nm, whereas the 1088-nm scheme achieves a detectable photon rate on the order of $10^3~\mathrm{s}^{-1}$. By quantifying the trade-off between irradiation time and scan-step size, we show that the nuclear transition can be located within one month for a 100-MHz uncertainty using currently available vacuum-ultraviolet laser technology. These results provide practical guidance for trapped-ion $^{229}\mathrm{Th}$ spectroscopy and the development of nuclear clocks.
\end{abstract}


\maketitle

\section{Introduction}
\begin{figure*}[t]
	\includegraphics[width=5.8in]{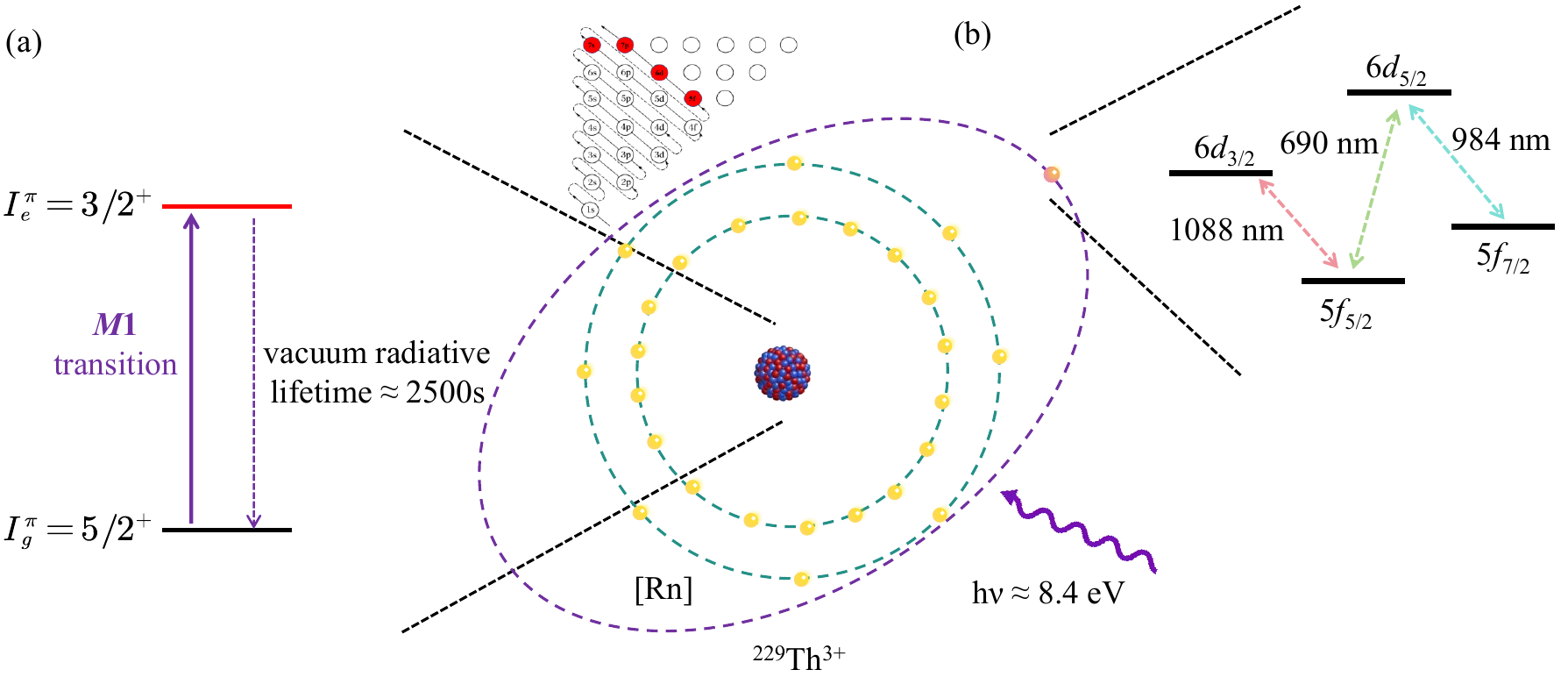}
	\caption{\label{IsomerLevel} (a) Nuclear ground and isomeric levels of $^{229}\mathrm{Th}$, connected by an $M1$ transition with energy $h\nu \approx 8.4\,\mathrm{eV}$. The vacuum radiative lifetime of the isomer is about $2500\,\mathrm{s}$. (b) Partial low-lying electronic levels of $^{229}\mathrm{Th}^{3+}$, with $5f_{5/2}$ as the electronic ground level. The 690-nm, 984-nm, and 1088-nm transitions are used in the proposed electronic-fluorescence detection schemes.
	}
\end{figure*}
The extremely low-energy nuclear isomeric state in $^{229}$Th, with an excitation energy of about 8~eV, is unique in its accessibility to laser spectroscopy. Controlled laser excitation of this isomer constitutes a crucial first step toward the development of nuclear optical clocks \cite{Peik2003,Rellergert2010,Campbell2012,Kazakov2012} and precision tests of fundamental physics \cite{Flambaum2006,Hayes2007,He2008,Fadeev2020,Peik2021,Wang2025,Beeks2025}. Current experimental efforts toward a $^{229}$Th nuclear clock are primarily focused on solid-state and trapped-ion platforms, where substantial progress has been made (see \cite{Lu2026Th229,Luo2026,Xu2026,Gan2026,Thirolf2024} for comprehensive reviews).

In 2024, by employing tunable 148-nm vacuum-ultraviolet (VUV) laser sources, the first resonant excitation of the $^{229}$Th isomer was realized in a $^{229}$Th-doped CaF$_2$ crystal \cite{Tiedau2024} and subsequently confirmed in $\mathrm{LiSrAlF}_6$ \cite{Elwell2024}. 
Shortly thereafter, a kilohertz-level precision measurement of the nuclear clock transition frequency was achieved in 
a CaF$_2$ host using a VUV frequency comb \cite{Zhang2024}. Parallel efforts have focused on understanding and controlling nuclear quenching mechanisms to optimize clock interrogation protocols and performance \cite{Schaden2025,Terhune2025}. Beyond bulk crystals, VUV excitation has also been reported in thin $^{229}$Th films and in ThO$_2$ samples \cite{Zhang2024b,Elwell2025}. 

Compared with solid-state systems, trapped ions provide a unique platform for a $^{229}$Th nuclear clock because of their excellent environmental control and well-characterized systematic shifts. In particular, $^{229}$Th$^{3+}$ is considered an especially promising candidate system. It has been extensively studied theoretically \cite{Peik2003,Campbell2012,Safronova2006,Porsev2010a,Safronova2013,Beloy2014,Li2021,Beloy2023,Wang2024,Zhou2026,Chakraborty2026}, and significant experimental progress has been achieved in ion production, trapping, cooling, and precision hyperfine spectroscopy \cite{Campbell2009,Campbell2011,Thielking2018,Scharl2023,Zitzer2024,Yamaguchi2024,Moritz2025,Zitzer2025,Li2025}.

However, direct laser excitation of the nuclear isomer has not yet been achieved in trapped-ion systems. The primary challenges arise from the extremely limited number of ions (typically tens to hundreds), the correspondingly weak nuclear fluorescence signal, and the several-hundred-megahertz uncertainty in the nuclear transition frequency \cite{Dzuba2023,Perera2025,Si2025}. Overcoming these limitations requires the development of higher-power VUV laser sources and more efficient nuclear-state detection methods. With the recent development of continuous-wave (CW) VUV laser based on four-wave mixing in cadmium vapor \cite{Xiao2026} and second-harmonic generation in SrB$_4$O$_7$ crystal \cite{Lal2025}, together with the aforementioned experimental progress, a systematic theoretical investigation of VUV laser excitation and detection schemes for the nuclear isomer in ion traps has become essential to support further experimental investigations.

In this work, we present a comprehensive quantum-optical investigation of hyperfine-resolved excitation and detection of the $^{229}\mathrm{Th}$ nuclear isomer in trapped $^{229}$Th$^{3+}$ ions.  We quantitatively analyze the direct nuclear excitation dynamics driven by experimentally available CW 148-nm VUV laser parameters, an aspect that has not been investigated in detail in previous trapped-ion studies. Going beyond earlier conceptual discussions of electronic-fluorescence-based nuclear-state detection, we further develop concrete detection schemes based on hyperfine-resolved fluorescence signals associated with the 690-, 984-, and 1088-nm electronic transitions, in which a specific laser-polarization-switching method is incorporated to prevent population trapping in uncoupled magnetic dark states. We also quantitatively estimate the corresponding photon count rates. On the basis of this quantitative treatment, we optimize the search strategy and provide estimates of the time required to locate the nuclear transition. Our results provide quantitative guidance for trapped-ion experiments and establish a theoretical foundation for the near-future realization of laser-driven nuclear excitation in this platform.

The paper is organized as follows. In Sec.~\ref{sec_2026_2_23_1}, we present the specific hyperfine levels of $^{229}\mathrm{Th}^{3+}$ and $^{229\mathrm{m}}\mathrm{Th}^{3+}$ involved in our schemes. 
In Secs.~\ref{sec_2026_3_17_1} and \ref{sec_2026_3_25_1}, we formulate hyperfine-resolved quantum-optical models to describe the laser-driven excitation and detection of the nuclear isomer. Finally, Sec.~\ref{sec_2026_2_22_1} presents a detailed analysis of the excitation and detection dynamics, and provides estimates for the detectable photon count and the total search time for the nuclear isomer.

\section{\label{sec_2026_2_23_1}Hyperfine levels in $^{229}\mathrm{Th}^{3+}$ and $^{229\mathrm{m}}\mathrm{Th}^{3+}$ }

\subsection{\label{sec_2026_2_23_2}General theory of hyperfine levels}

In atoms, the hyperfine interaction couples the nuclear and electronic degrees of freedom.
As a consequence, the nuclear spin $I$ and the electronic angular momentum $J$ are no longer individually conserved. 
Instead, the total angular momentum $F$, arising from the coupling of $I$ and $J$ (where both are non-zero), emerges 
as the new good quantum number of the coupled system. This coupling also leads to the 
splitting of an atomic level ($I$, $\Gamma J$) into multiple hyperfine sub-levels ($I$, $\Gamma J$, $F$), where $\Gamma$ denotes all additional electronic quantum numbers. Each hyperfine level ($I$, $\Gamma J$, $F$) is described by a dressed hyperfine state $\ket{[I\Gamma J] Fm}$  \cite{Wang2024,Wang2025,Wang2025a}, characterized by the same set of quantum numbers, where $m$ is the projection of $F$ along the quantization axis.
\begin{figure*}[t]
	\includegraphics[width=6.2in]{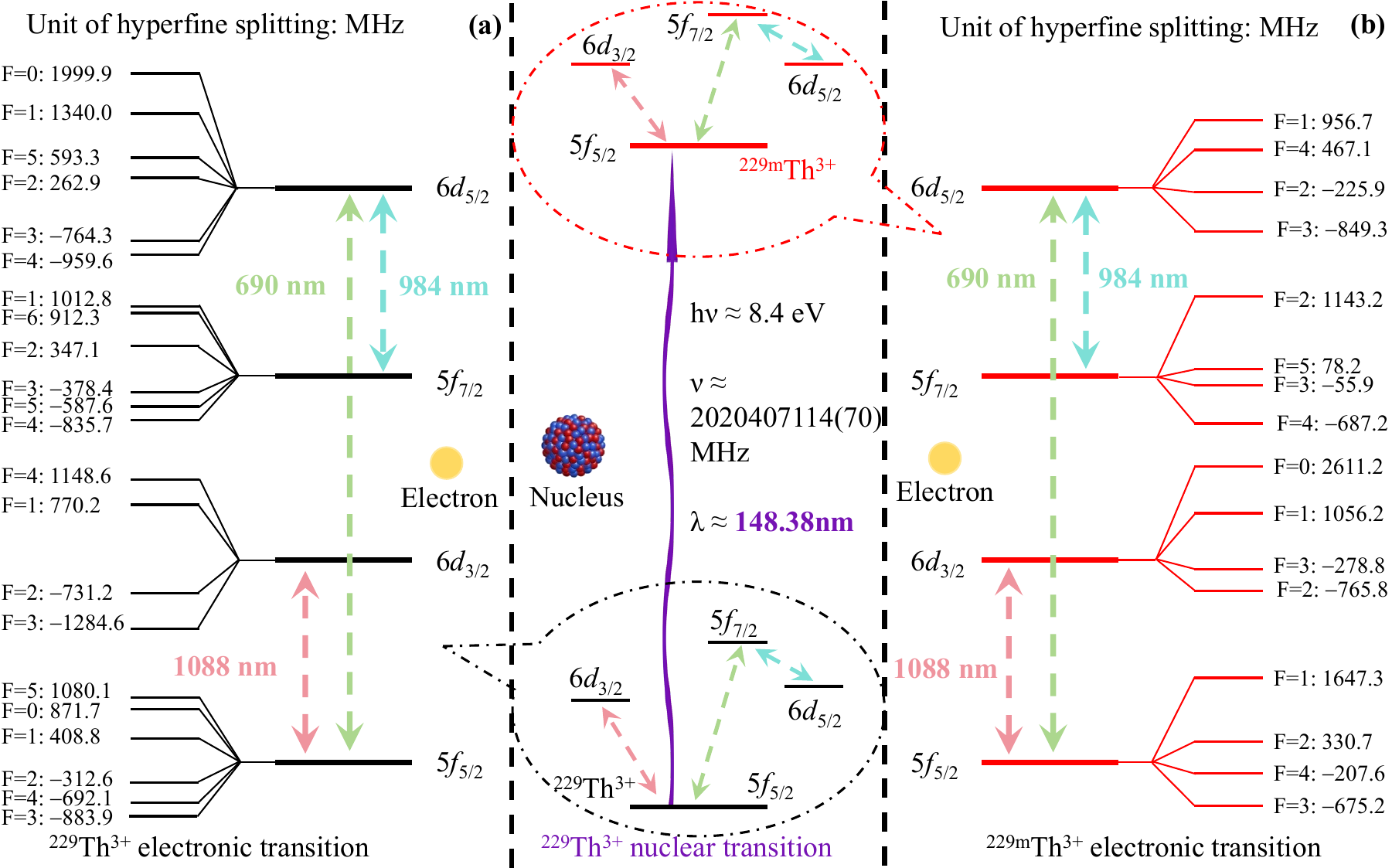}
	\caption{\label{HyperfineLevel} 
		Hyperfine structures of the selected electronic $5f_{5/2}$, $5f_{7/2}$, $6d_{3/2}$, and $6d_{5/2}$ levels in (a) $^{229}\mathrm{Th}^{3+}$ and (b) $^{229\mathrm{m}}\mathrm{Th}^{3+}$. The hyperfine splittings are given in units of $\mathrm{MHz}$. Hyperfine-resolved laser excitation is essential for improving the precision of the isomeric energy,
		while the distinct hyperfine structures of the nuclear ground and isomeric states enable nuclear-state detection via hyperfine-resolved electronic fluorescence.
	}
\end{figure*}

The dressed hyperfine state $\ket{[I\Gamma J] Fm}$ serves as 
an eigenstate of the coupled nucleus-electron system and accounts for nuclear hyperfine 
mixing  \cite{Wang2024,Wang2025,Wang2025a}. In this work, the effect of such mixing in $^{229}\mathrm{Th}^{3+}$ ions on the laser excitation and detection dynamics is negligible. Therefore, in the following calculations, the dressed hyperfine state $\ket{[I\Gamma J] Fm}$ is approximated by its leading term $\ket{I\Gamma J;Fm}$.
Here, $\ket{I\Gamma J;Fm}$ denotes the hyperfine-coupled basis state formed by the angular-momentum coupling of the corresponding nuclear and electronic states \cite{Fritzsche2012}.

The hyperfine interaction induces energy shifts of the hyperfine level ($I$, $\Gamma J$, $F$) relative to the unperturbed atomic level ($I$, $\Gamma J$). In first-order perturbation theory, the shifts are determined by the expectation values of the magnetic-dipole ($M1$) and electric-quadrupole ($E2$) components of hyperfine interaction, evaluated in the hyperfine state $\ket{I\Gamma J;Fm}$ (see, e.g., Ref.~\cite{Robert2018}): 
\begin{equation}\label{2026_2_23_1}
	\begin{split}
		E_{M1}&=\dfrac{1}{2} AK,\\
	E_{E2}&=\dfrac{1}{2} B\dfrac{3K(K+1)-4I(I+1)J(J+1)}{2I(2I-1)2J(2J-1)},
	\end{split}
\end{equation}
where $K=F(F+1)-J(J+1)-I(I+1)$, while $A$ and $B$ are hyperfine constants, respectively, defined by
\begin{equation}\label{2026_2_24_1}
	\begin{split}
		A&=\dfrac{\mu}{I}\dfrac{1}{\sqrt{J(J+1)(2J+1)}}\melb{\Gamma J}{\big|T^{(M1)}\big|}{\Gamma J},\\
		B&=2Q\sqrt{\dfrac{J(2J-1)}{(J+1)(2J+1)(2J+3)}}\melb{\Gamma J}{\big|T^{(E2)}\big|}{\Gamma J}.
	\end{split}
\end{equation}
Here, $\mu$ and $Q$ denote the magnetic dipole and electric quadrupole moments of the nuclear level with spin $I$, respectively. $T^{(M1)}$ and $T^{(E2)}$ represent the electronic $M1$ and $E2$ operators associated with the hyperfine interaction \cite{Wang2025a,Wang2023}.

\subsection{ \label{sec_2026_2_25_1}$^{229}\mathrm{Th}^{3+}$ and $^{229\mathrm{m}}\mathrm{Th}^{3+}$ ions}

The $^{229}\mathrm{Th}$ isomeric level with spin-parity $I_e^\pi=3/2^+$ is connected to its ground level with spin-parity $I_g^\pi=5/2^+$ via an $M1$ transition, as illustrated in Fig.~\ref{IsomerLevel}(a). This isomer possesses a vacuum radiative lifetime of about 2500 s \cite{Tiedau2024}. Due to the distinct nuclear spins as well as nuclear magnetic dipole and electric quadrupole moments of the ground and isomeric levels (see Table \ref{tab_2026_2_24_1}), their respective hyperfine levels exhibit significantly different structures and energy splittings.

\begin{table}[t]
	\caption{Nuclear magnetic dipole and electric quadrupole moments of $^{229}\mathrm{Th}$ and $^{229\mathrm{m}}\mathrm{Th}$, where $\mu_N$ denotes the nuclear magneton and eb represents the electron-barn. }\label{tab_2026_2_24_1}
	\vspace{0.7em}\centering
	\setlength{\tabcolsep}{8pt}
	\renewcommand{\arraystretch}{1.4}
	\begin{tabular}{cccccccccc}
		\hline
		\hline
		Nucleus \  & \ $\mu~(\mu_N)$ \  & \  $Q$ (eb)\ & Reference  \\
		\hline
		$^{229}\mathrm{Th}$  \ & \ 0.365(3)  \  &  \ 3.11(5)  \  &   Ref.~\cite{Zitzer2025}\\
		$^{229\mathrm{m}}\mathrm{Th}$	\	 & \ $-$0.378(8)   \  & \ 1.77(2)   \ &    Ref.~\cite{Yamaguchi2024} \\
		\hline
		\hline
	\end{tabular}
\end{table}

The electronic configuration of the Th$^{3+}$ ion consists of a closed [Rn] core and a single valence electron.
For brevity, we denote electronic levels in Th$^{3+}$ solely by the valence-electron configurations throughout the text.
For the electronic levels considered in this work [see Fig.~\ref{IsomerLevel}(b)], the corresponding hyperfine constants---$A_g$ and $B_g$ for $^{229}\mathrm{Th}^{3+}$, and $A_e$ and $B_e$ for $^{229\mathrm{m}}\mathrm{Th}^{3+}$---are summarized in Table~\ref{tab_2026_2_25_1}.  The constants $A_g$ and $B_g$ are taken from experimental measurements, whereas experimental data for $A_e$ and $B_e$ are available only for the electronic $6d_{3/2}$ level. The remaining values are determined using the following formulae derived from Eq.~\eqref{2026_2_24_1}: 
\begin{equation}\label{2026_2_24_2}
	A_e=A_g\dfrac{I_g\mu_e}{I_e\mu_g} ,~~~
B_e=B_g\dfrac{Q_e}{Q_g}.
\end{equation}

Figure~\ref{HyperfineLevel} illustrates the partial hyperfine levels of $^{229}\mathrm{Th}^{3+}$ and $^{229\mathrm{m}}\mathrm{Th}^{3+}$, with the corresponding energy splittings calculated using Eq.~\eqref{2026_2_23_1}. In these hyperfine levels, their energy splittings range from several hundred MHz to several GHz. For a given electronic level, the difference in hyperfine splitting between $^{229}\mathrm{Th}^{3+}$ and $^{229\mathrm{m}}\mathrm{Th}^{3+}$ is on the order of several hundred MHz.

In the laser excitation of $^{229\mathrm{m}}\mathrm{Th}^{3+}$, a narrow-linewidth laser is essential, both to achieve sub-MHz isomeric energy measurement precision and to ensure high excitation efficiency. Consequently, a hyperfine-resolved laser excitation scheme is required and investigated in this work. 

Direct nuclear fluorescence detection of the isomer is highly inefficient because of its long radiative lifetime and the limited number of trapped $^{229\mathrm{m}}\mathrm{Th}^{3+}$ ions. Recent experiments using CW nuclear laser absorption spectroscopy in $^{229}\mathrm{Th}$-doped crystals have demonstrated that direct absorption provides an effective nuclear-state readout for systems containing large ensembles of nuclei \cite{Morawetz2026CWAbsorption,ToscaniDeCol2026ThoriumClock,Huang2026NuclearClock}. However, in trapped-ion experiments, where only tens of ions are typically available, the nuclear absorption signal is intrinsically weak and difficult to distinguish from background noise. In contrast, the distinct hyperfine levels of $^{229}\mathrm{Th}^{3+}$ and $^{229\mathrm{m}}\mathrm{Th}^{3+}$ enable an alternative detection scheme: by probing selected electronic transitions with a hyperfine-resolved laser, the nuclear level can be identified via the enhancement or suppression of the electronic resonance fluorescence \cite{Peik2003}. This method is considerably more efficient due to the microsecond-scale lifetimes of the electronic excited states \cite{Safronova2006}.

Given these considerations, we focus on hyperfine-resolved schemes for both nuclear laser excitation and detection in the following sections.
\begin{table}[t!]
	\caption{Hyperfine constants $A_g$ ($A_e$) and $B_g$ ($B_e$) for partial electronic levels of $^{229}\mathrm{Th}^{3+}$ ($^{229\mathrm{m}}\mathrm{Th}^{3+}$). }\label{tab_2026_2_25_1}
	\vspace{0.7em}\centering
	\setlength{\tabcolsep}{4.5pt}
	\renewcommand{\arraystretch}{1.4}
	\begin{tabular}{ccccc}
		\hline
		\hline
		Level & $A_g$    (MHz)&   $B_g$ (MHz)& $A_e$   (MHz) & $B_e$  (MHz) \\
		\hline
		$5f_{5/2}$    & 82.0(2)$^a$        & 2270.3(18)$^a$       & $ -141.5(32)$        & 1292(25) \\
		$5f_{7/2}$ 	  & 31.4(7)$^b$          & 2550(12)$^b$          &$-54.2(17)$           &   1451(29)\\
		$6d_{3/2}$ 	 &  155.3(12)$^b$     & 2265(9)$^b$           &   $-$267(3)$^c$  &1288(10)$^c$  \\
		$6d_{5/2}$  &  $-$12.9(3)$^a$   & 2695.7(19)$^a$    &   $22.3(7)$  			& 1534(30) \\
		\hline
		\hline
	\end{tabular}
	\raggedright
	$^a$Ref.~\cite{Zitzer2025}; $^b$Ref.~\cite{Campbell2011}; $^c$Ref.~\cite{Yamaguchi2024}
\end{table}
\section{\label{sec_2026_3_17_1}Theory of direct nuclear laser excitation in $^{229}\mathrm{Th}^{3+}$}
\begin{figure}[t]
	\includegraphics[width=3.3in]{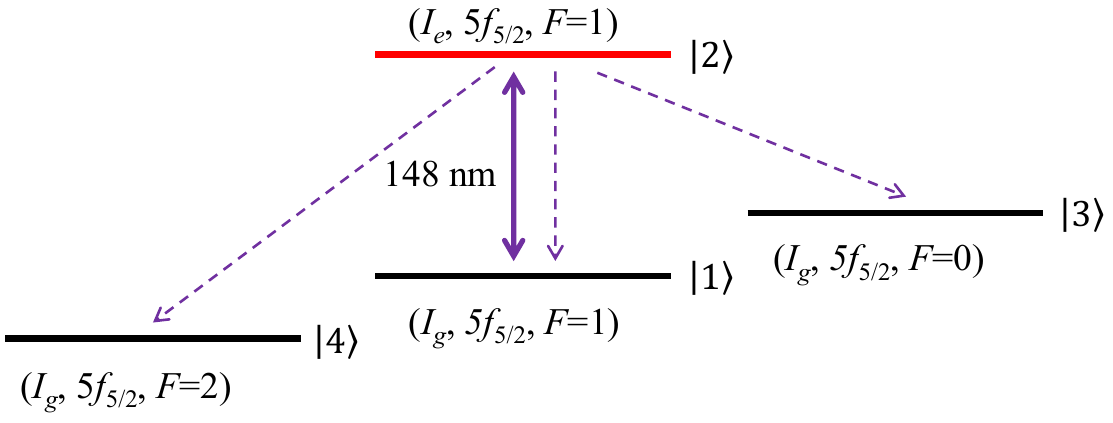}
	\caption{\label{ExcitationDynamics} Four-level model for direct laser excitation of the nuclear isomer in $^{229}\mathrm{Th}^{3+}$. The 148-nm laser drives the transition between states $|1\rangle$ and $|2\rangle$, which belong to the hyperfine levels $(I_g,5f_{5/2},F=1)$ and $(I_e,5f_{5/2},F=1)$, respectively. The excited state $|2\rangle$ can decay to $|1\rangle$ as well as to $|3\rangle$ and $|4\rangle$, which belong to the hyperfine levels  $(I_g, 5f_{5/2}, F=0)$ and $(I_g, 5f_{5/2}, F=2)$, respectively.
	}
\end{figure}

In this section, we investigate the direct optical excitation of the $^{229}\mathrm{Th}$ isomer using a 148-nm VUV laser. The system is assumed to be initially prepared in the nuclear ground level $(I_g, 5f_{5/2}, F=1)$ via optical pumping (see Appendix~\ref{app_initial_state}) and subsequently driven to the nuclear isomeric level $(I_e, 5f_{5/2}, F=1)$ by the laser field. The isomeric level $(I_e, 5f_{5/2}, F=1)$ can decay into three hyperfine levels $(I_g, 5f_{5/2}, F=0, 1, 2)$. To describe the excitation dynamics, we employ a four-level model, as illustrated in Fig.~\ref{ExcitationDynamics}. As discussed in Sec.~\ref{sec_2026_2_23_2},  these four hyperfine levels are represented by the hyperfine states $\ket{I_g5f_{5/2};F=1m_1}$  ($\equiv \ket{1}$), $\ket{I_e5f_{5/2};F=1m_2}$ ($\equiv \ket{2}$), $\ket{I_g5f_{5/2};F=0m_3}$ ($\equiv \ket{3}$), and $\ket{I_g5f_{5/2};F=2m_4}$ ($\equiv \ket{4}$), where $m_i$ denotes the magnetic quantum number. Atomic units ($\hbar = |e| = m_e = 1$ and $c = 1/\alpha$) are used throughout this work.

\subsection{Quantum master equation for nuclear transition}
The time evolution of the four-level system is described by the quantum master equation \cite{Budini2000,Kiely2021}
\begin{equation}\label{2025_11_29_7}
	\dot{\rho}=-i[H_I, \rho]+\dfrac{\gamma_{n,l}}{2}\mathcal{L}_{\hat{\sigma}_{22}}\rho+\mathcal{L}\rho,
\end{equation}
where the Hamiltonian $H_I$ is  given by
\begin{equation}\label{2025_12_4_1}
	H_I=\Delta_{n,l}\hat{\sigma}_{22}
	+[\Omega_n(t)\hat{\sigma}_{21}+\mathrm{H.c.}]
\end{equation}
in the interaction picture. Here, $\Delta_{n,l} = \omega_n - \omega_l$ denotes the detuning between the isomeric energy $\omega_n$ and the laser frequency $\omega_l$, and $\Omega_n(t)$ is  the nuclear Rabi frequency. The parameter $\gamma_{n,l}$ represents the laser linewidth \cite{vonderWense2020,vonderWense2020b}, while $\mathcal{L}\rho$ is the Lindblad term given by 
\begin{equation}\label{2026_3_17_1}
\mathcal{L}\rho	=\frac{\gamma_{n,21}}{2} \mathcal{L}_{\hat{\sigma}_{12}}\rho+\frac{\gamma_{n,23}}{2} \mathcal{L}_{\hat{\sigma}_{32}}\rho+\frac{\gamma_{n,24}}{2} \mathcal{L}_{\hat{\sigma}_{42}}\rho,
\end{equation}
where $\gamma_{n,ij}$ is the decay rate from $\ket{i}$ to $\ket{j}$ and $\hat{\sigma}_{ij}=|i\rangle\langle j|$. The superoperator $\mathcal{L}_{\hat{\sigma}_{ji}}$ is defined as \cite{Scully1997}
\begin{equation}\label{2026_3_17_2}
	\mathcal{L}_{\hat{\sigma}_{ji}} \rho = 2\hat{\sigma}_{ji} \rho \hat{\sigma}_{ji}^\dagger - \hat{\sigma}_{ji}^\dagger \hat{\sigma}_{ji} \rho - \rho \hat{\sigma}_{ji}^\dagger \hat{\sigma}_{ji} .
\end{equation}

Under the rotating-wave approximation (RWA), the Rabi frequency $\Omega_n(t)$ is written as 
\begin{equation}\label{2025_12_6_1}
	\begin{split}
	\Omega_n(t)= &-\frac{F_{n,0}}{2\sqrt{2}}f_n(t)(-1)^{1-m_2}\bigg[\left(\begin{matrix}
		1     & 1     & 1\\
		-m_2& 1& m_1 \end{matrix}\right)- \\
		& 	\left(\begin{matrix}
			1     & 1     & 1\\
			-m_2& -1& m_1 \end{matrix}\right)\bigg]M_n,
	\end{split}
\end{equation}
where $F_{n,0}$ and $f_n(t)$ denote the amplitude and envelope function of the laser electric field, respectively, and $M_n$ denotes the reduced matrix element for the nuclear $M1$ transition, as defined in Eq.~\eqref{2025_12_5_1}.  The laser field is assumed to be linearly polarized along the $z$ axis.
Since the nonzero absolute values of $\Omega_n(t)$ are identical for different $m_1$, we choose $m_1 = 1$ and $m_2 = 0$ in the following calculations without loss of generality. Equation~\eqref{2025_12_6_1} is derived using the multipole expansion of the nucleus-laser interaction \cite{Wang2023}. A detailed derivation is provided in Appendix~\ref{app_2026_3_18_1}. 

The general expression for the nuclear decay rate $\gamma_n$ from the levels  ($I_e$, $\Gamma J$, $F_i$) to ($I_g$, $\Gamma J$, $F_f$) 
is 
\begin{equation}\label{2026_3_19_7}
\gamma_n=  \frac{4}{3}k_n^3[F_f] \left\{\begin{matrix}
	I_g& I_e & 1 \\
	F_i & F_f & J
\end{matrix}\right\}^2  |\melb{I_e }{\big|\mathcal{M}^{(M1)}\big|}{I_g }|^2,
\end{equation}
where $[F_f]\equiv2F_f+1$, $k_n=\omega_n/c$, and the reduced matrix element $\melb{I_e }{\big|\mathcal{M}^{(M1)}\big|}{I_g }$ is determined by the reduced nuclear transition probability $B(M1,I_e\rightarrow I_g)$ [see Eq.~\eqref{2026_3_19_8}].
In this work, $B(M1,I_e \rightarrow I_g)=0.022$ W.u. is adopted \cite{Tiedau2024}. Using Eq.~\eqref{2026_3_19_7}, we obtain $\gamma_{n,21}/2\pi=3.0\times10^{-5}$~Hz, $\gamma_{n,23}/2\pi=1.4\times10^{-5}$~Hz, and
 $\gamma_{n,24}/2\pi=2.0\times10^{-5}$~Hz.

\subsection{Quasi-steady state in nuclear excitation}

\begin{figure*}[t]
	\includegraphics[width=5.6 in]{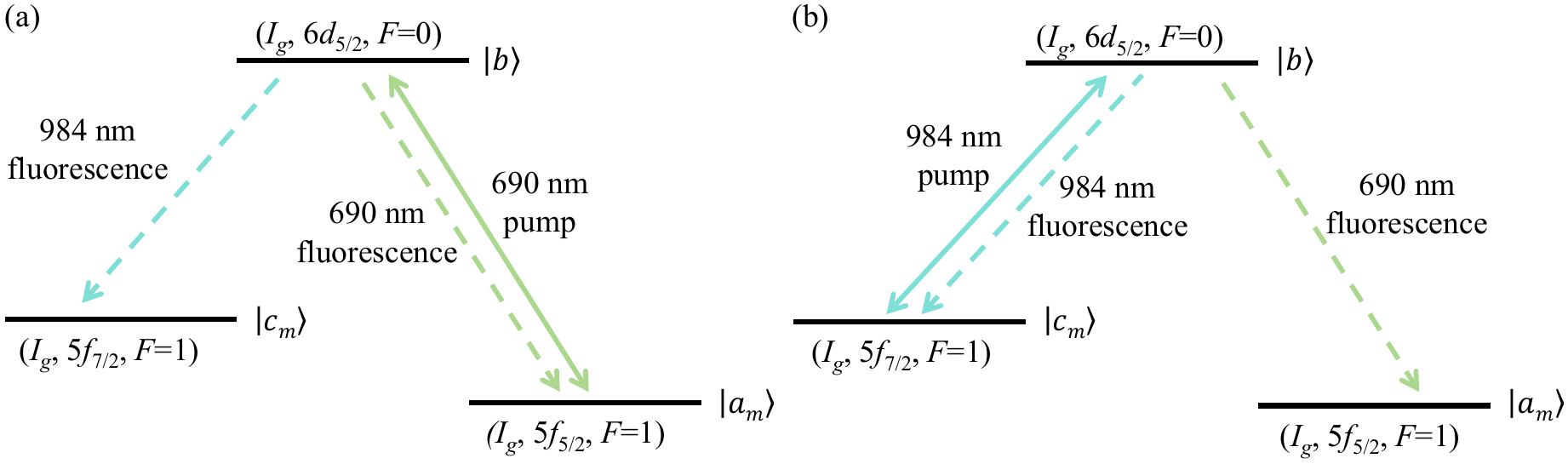}
	\caption{\label{Detection690nm984nm} 
		Seven-level detection scheme based on the 690-nm and 984-nm electronic transitions in the nuclear-ground-state configuration. (a) During the 690-nm pumping step, the electronic population is driven from states $\ket{a_m}$ to $\ket{c_m}$ ($m=0,\pm1$) via the intermediate state $\ket{b}$, accompanied by the emission of 690- and 984-nm fluorescence. The states $\ket{a_m}$, $\ket{b}$, and $\ket{c_m}$ belong to the hyperfine levels $(I_g, 5f_{5/2}, F=1)$, $(I_g, 6d_{5/2}, F=0)$, and $(I_g, 5f_{7/2}, F=1)$, respectively. (b) During the 984-nm pumping step, conversely, the electronic population is driven from $\ket{c_m}$ back to $\ket{a_m}$ via $\ket{b}$, resulting in the emission of 690- and 984-nm  fluorescence. The 690-nm or 984-nm  fluorescence is suppressed when the nuclear isomer is excited.
	}
\end{figure*}
For short times (e.g., $\lesssim$ 500 s), the decay of the nuclear isomer can be neglected, and Eq.~\eqref{2025_11_29_7} reduces to a two-level quantum master equation,
\begin{equation}\label{2026_3_19_9}
	\dot{\rho}=-i[H_I, \rho]+\dfrac{\gamma_{n,l}}{2}\mathcal{L}_{\hat{\sigma}_{22}}\rho.
\end{equation}
In the presence of the laser linewidth ($\gamma_{n,l}\neq0$), the above equation admits an exact solution at resonance ($\Delta_{n,l}=0$), known as Torrey's solution \cite{Torrey1949,Steck2026}. For the case of nonzero detuning ($\Delta_{n,l}\neq0$), however, no simple analytic solution exists. Nevertheless, in the regime where the absolute value of the nuclear Rabi frequency is much smaller than the laser linewidth ($|\Omega_n|\ll\gamma_{n,l}$), the two-level system can be treated using the adiabatic approximation \cite{Steck2026}. 

The component form of Eq.~\eqref{2026_3_19_9} is explicitly given by \cite{vonderWense2020}
\begin{equation}\label{2025_12_8_1}
	\begin{split}
		\dot{\rho}_{22}=& -i\Omega_n(	\rho_{12}-	\rho_{21}),\\
		\dot{\rho}_{12}=	&-i\Omega_n(	\rho_{22}-	\rho_{11})+i\Delta_{n,l}	\rho_{12}-\frac{\gamma_{n,l}}{2}\rho_{12},
	\end{split}
\end{equation}
where the envelope function is taken to be constant [$f(t)=1$]. Note that $\rho_{21}=\rho_{12}^*$ and $\rho_{11}+\rho_{22}=1$.
By applying the adiabatic approximation $\dot{\rho}_{12}\approx0$, the second equation of Eq.~\eqref{2025_12_8_1} yields
\begin{equation}\label{2025_12_8_2}
	\rho_{12}=\dfrac{2i\Omega_n(\rho_{11}-\rho_{22})}{\gamma_{n,l}-2i\Delta_{n,l}	}.
\end{equation}
Substituting Eq.~\eqref{2025_12_8_2} into the first equation of Eq.~\eqref{2025_12_8_1} and using the initial condition $\rho_{22}(0)=0$, we obtain
\begin{equation}\label{2025_12_8_4}
\rho_{22}(t)=\dfrac{1}{2}\left[1-\mathrm{exp}\left( -\dfrac{8\gamma_{n,l} \Omega_n^2 }{4\Delta_{n,l}^2+\gamma_{n,l}^2}t\right)\right].
\end{equation}

It follows from Eq.~\eqref{2025_12_8_4} that the two-level system can reach a quasi-steady state where the isomeric population saturates at $1/2$ in the regime of large laser linewidth. The characteristic time required to achieve this quasi-steady state is
\begin{equation}\label{2025_12_8_5}
	T_0=\dfrac{4\Delta_{n,l}^2+\gamma_{n,l}^2 }{ 8\gamma_{n,l} \Omega_n^2 }.
\end{equation}
This characteristic time indicates that, when the laser detuning is smaller than the laser linewidth ($\Delta_{n,l}\lesssim\gamma_{n,l}$), the time required to reach the quasi-steady state increases with the laser linewidth. For a fixed laser linewidth, an increase in laser detuning also extends the time to achieve this state.

\section{\label{sec_2026_3_25_1} Detection schemes of $^{229}\mathrm{Th}$ isomer via electronic fluorescence}

As discussed at the end of Sec.~\ref{sec_2026_2_25_1}, indirect detection of the nuclear level via the electronic fluorescence is more efficient than direct detection through nuclear fluorescence. In this section, we present the theoretical framework for detection schemes based on electronic fluorescence.

\subsection{\label{sec_2026_7_13_1}  Detection scheme using 690-nm and 984-nm lasers }

One possible scheme is to use the 690-nm and 984-nm lasers to drive electronic transitions between the $5f_{5/2}$ and $5f_{7/2}$ levels via the intermediate $6d_{5/2}$ level.  As illustrated in Fig.~\ref{Detection690nm984nm}(a), when the 690-nm laser is applied, the electronic population is transferred from $5f_{5/2}$ to $5f_{7/2}$ levels, accompanied by the emission of 690- and 984-nm fluorescence. Conversely, as shown in Fig.~\ref{Detection690nm984nm}(b), when the 984-nm laser is applied, the  electronic population is transferred from $5f_{7/2}$ to $5f_{5/2}$ levels, accompanied by the emission of 690- and 984-nm  fluorescence. Consequently, these two processes provide different detection methods based on the observation of either 690-nm or 984-nm fluorescence.

As shown in Table~\ref{tab_2026_2_25_1}, the hyperfine levels in $^{229}\mathrm{Th}^{3+}$ are more precisely known than those in $^{229\mathrm{m}}\mathrm{Th}^{3+}$. Therefore, we construct our scheme based on the hyperfine levels in $^{229}\mathrm{Th}^{3+}$.
Specifically, we consider a seven-level system composed of the hyperfine levels $(I_g, 5f_{5/2}, F=1)$, $(I_g, 6d_{5/2}, F=0)$, and $(I_g, 5f_{7/2}, F=1)$, as shown in Fig.~\ref{Detection690nm984nm}.
Since the two levels with $F=1$ contain degenerate magnetic sublevels, we explicitly write $\ket{a_m}
	\equiv \ket{I_g5f_{5/2};F=1m}, ~m=-1,0,1$, $\ket{b} \equiv \ket{I_g6d_{5/2};F=0m=0}$, $\ket{c_m}
	\equiv \ket{I_g5f_{7/2};F=1m},~ m=-1,0,1.$ In this scheme, if the nuclear isomer is excited, the electronic fluorescence signal is significantly suppressed.

The dynamics of the even-level system obeys the quantum master equation
\begin{equation}\label{2026_3_20_1}
	\dot{\rho}=-i[H_I^{(q)}, \rho]+\dfrac{\gamma_{e,l}}{2}\mathcal{L}_{\hat{\sigma}_{bb}}\rho+\mathcal{L}\rho,
\end{equation}
where $\gamma_{e,l}$ denotes the linewidth of the 690-nm or 984-nm laser, and $H_I^{(q)}$ is given, in the interaction picture, by
\begin{equation}\label{2026_3_21_1}
	H_I^{(q)}=\Delta_{e,l}\hat{\sigma}_{bb}
	+[\Omega_e^{(q)}(t)\hat{\sigma}_{b\alpha}+\mathrm{H.c.}]
\end{equation}
Here, $\Delta_{e,l}$ is the detuning between the electronic transition energy and the frequency of the 690-nm or 984-nm laser. The operator
$\hat{\sigma}_{b\alpha}$ represents $\hat{\sigma}_{ba_{m_i}}$ for the 690-nm transition and $\hat{\sigma}_{bc_{m_f}}$ for the 984-nm transition. The index $q$ labels the polarization component of the laser field, with $q=0$ corresponding to linear polarization along the $z$ axis and $q=\pm1$ corresponding to the two circular polarizations. The  Lindblad term
$\mathcal{L}\rho$ is expressed as
\begin{equation}\label{2026_3_21_2}
	\mathcal{L}\rho	=\sum_{m=-1}^{1}\bigg(
	\frac{\gamma_{e,ba_m}}{2}
	\mathcal{L}_{\hat{\sigma}_{a_m b}}\rho
	+
	\frac{\gamma_{e,bc_m}}{2}
	\mathcal{L}_{\hat{\sigma}_{c_m b}}\rho\bigg) ,
\end{equation}
where $\gamma_{e,ba_m}$ and $\gamma_{e,bc_m}$ are the decay rates from states $\ket{b}$ to $\ket{a_m}$
and $\ket{c_m}$, respectively.

By applying RWA, the electronic Rabi frequency $\Omega_e^{(q)}(t)$ for the transition from the hyperfine states $\ket{I\Gamma_i J_i;F_im_i}$ to  $\ket{I\Gamma_fJ_f;F_fm_f}$ is calculated to be 
\begin{equation}\label{2025_12_6_4}
\Omega_e^{(q)}(t)=\frac{F_{e,0}}{2}f_e(t)(-1)^{F_f-m_f}\left(\begin{matrix}
		F_f      & 1     &  F_i \\
		-m_f& q& m_i\end{matrix}\right)M_e,
\end{equation}
where $F_{e,0}$ and $f_e(t)$ are the amplitude and envelope function of the electric field of the 690-nm or 984-nm laser, respectively, and $M_e$ denotes the reduced matrix element of the electronic $E1$ transition given by Eq.~\eqref{2026_3_22_1}. The detailed derivation of Eq.~\eqref{2025_12_6_4} can be found in Refs.~\cite{Wang2024,Wang2026}.

For the 690- and 984-nm transitions considered here, the lower level has $F_i=1$, whereas the upper state has $F_f=0$ and $m_f=0$. Conservation of angular momentum imposes the selection rule $m_i+q=0$. Consequently, for a laser beam with fixed polarization, only one magnetic sublevel is coupled to $\ket{b}$, while the other two magnetic sublevels in the $\ket{a_m}$ or $\ket{c_m}$ manifold remain uncoupled and thus behave as dark states. To avoid population trapping in these dark sublevels, the laser polarization is alternated among the $q=0$, $q=+1$, and $q=-1$ components, so that all three magnetic sublevels are driven sequentially. The detailed scheme is described in Sec.~\ref{sec_2026_7_13_2}.

The general expression for the electronic decay rate $\gamma_e$ from the levels $(I,\Gamma_i J_i,F_i)$ to $(I,\Gamma_f J_f,F_f)$ is 
written as
\begin{equation}\label{2025_12_6_6}
	\gamma_e=  \frac{4}{3}k_e^3[F_f] \left\{\begin{matrix}
		J_f& J_i & 1 \\
		F_i & F_f & I
	\end{matrix}\right\}^2  |\melb{\Gamma_{f}J_{f}}{\big|\mathcal{O}^{(E1)}\big|}{\Gamma_{i}J_{i}}|^2,
\end{equation}
where $k_e=\omega_e/c$ with $\omega_e$ being the electronic transition energy. Equation~\eqref{2025_12_6_6} gives the electronic decay rate summed over the magnetic sublevels. To describe the magnetic-sublevel-resolved decay channels, we introduce the decay rate from $\ket{I\Gamma_i J_i;F_im_i}$ to $\ket{I\Gamma_fJ_f;F_fm_f}$, which is given by
\begin{equation}\label{2026_7_8_1}
	\gamma_{e}(m_i\rightarrow m_f)=  [F_i] \left|
	\begin{pmatrix}
		F_f & 1 & F_i \\
		-m_f & m_f-m_i& m_i
	\end{pmatrix}
	\right|^2 \gamma_e.
\end{equation}
The relation between the reduced matrix element $\melb{\Gamma_{f}J_{f}}{\big|\mathcal{O}^{(E1)}\big|}{\Gamma_{i}J_{i}}$ and the lifetime of the electronic level is given in Appendix~\ref{app_2026_3_22_1}.
The lifetimes of the transitions $6d_{5/2}\rightarrow5f_{5/2}$ and $6d_{5/2}\rightarrow5f_{7/2}$ are 5.8 ~$\mu$s and 0.8~$\mu$s \cite{Safronova2006}, respectively. In the present case, using Eq.~\eqref{2026_7_8_1}, we obtain the magnetic-sublevel-resolved decay rates $\gamma_{e,ba_m}/2\pi= 9.2\times10^3$~Hz and $\gamma_{e,bc_m}/2\pi= 6.9\times10^4$~Hz for $m=0,\pm1$.

\subsection{Detection scheme using the 1088-nm  laser}

\begin{figure}[t]
	\includegraphics[width=1.4in]{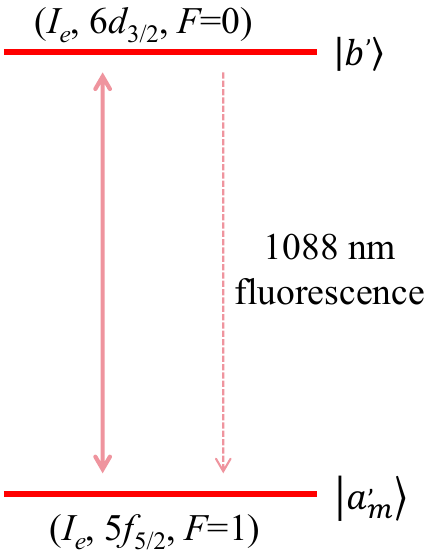}
	\caption{\label{Detection1088nm}
		Four-level detection scheme based on the 1088-nm electronic transition in the nuclear-isomeric-state configuration. The laser couples the states $\ket{a^\prime_m}$ ($m=0,\pm1$)  and $\ket{b^\prime}$, which belong to the
		hyperfine levels $(I_e,5f_{5/2},F=1)$ and $(I_e,6d_{3/2},F=0)$, respectively. The 1088-nm fluorescence is enhanced when the nuclear isomer is excited.
	}
\end{figure}

Another possible detection scheme employs a 1088-nm laser to drive the electronic transition between the $5f_{5/2}$ and $6d_{3/2}$ levels, thereby generating 1088-nm fluorescence. We consider the transition between the hyperfine levels $(I_e,5f_{5/2},F=1)$ and $(I_e,6d_{3/2},F=0)$, as shown in Fig.~\ref{Detection1088nm}. Because the lower hyperfine level has $F=1$, it contains three degenerate magnetic sublevels. We therefore define $\ket{a'_m}\equiv \ket{I_e5f_{5/2};F=1m}, \qquad m=-1,0,+1,$ and $\ket{b'}\equiv \ket{I_e6d_{3/2};F=0m=0}$. Since $(I_e,6d_{3/2},F=0)$ decays dominantly to $(I_e,5f_{5/2},F=1)$,
the detection dynamics can be described using a four-level model. As these levels belong to the nuclear isomer, the electronic fluorescence signal is significantly enhanced when the isomer is successfully excited.

The dynamics of the four-level system are governed by the quantum master equation
\begin{equation}\label{2026_7_22_1}
	\dot{\rho}
	=
	-i[H_I^{(q)},\rho]
	+
	\frac{\gamma^\prime_{e,l}}{2}
	\mathcal{L}_{\hat{\sigma}_{b'b'}}\rho
	+	\mathcal{L}\rho,
\end{equation}
where $\gamma^\prime_{e,l}$ denotes the linewidth of the 1088-nm laser, the interaction Hamiltonian is given by
\begin{equation}\label{2026_7_22_3}
	H_I^{(q)}=\Delta_{e,l}^\prime\hat{\sigma}_{b^\prime b^\prime}
	+[\Omega_e^{(q)}(t)\hat{\sigma}_{b^\prime a^\prime_m}+\mathrm{H.c.}],
\end{equation}
and the Lindblad term is
\begin{equation}\label{2026_7_22_2}
	\mathcal{L}\rho	=
\sum_{m=-1}^{1}
\frac{\gamma_{e,b^\prime a^\prime_m}}{2}
\mathcal{L}_{\hat{\sigma}_{a^\prime_m b^\prime}}\rho.
\end{equation}
Here, $\Delta_{e,l}^{\prime}$ denotes the detuning of the 1088-nm laser.
	The Rabi frequency $\Omega_e^{(q)}(t)$ is calculated using
	Eq.~\eqref{2025_12_6_4}, whereas the magnetic-sublevel-resolved decay rate
	$\gamma_{e,b^\prime a^\prime_m}$ is obtained from
	Eq.~\eqref{2026_7_8_1}.
	Using the approximately
	$1.1~\mu\mathrm{s}$ lifetime of the $6d_{3/2}$ level
	\cite{Safronova2006}, we obtain
	$\gamma_{e,b^\prime a^\prime_m}/2\pi
	=4.9\times10^4~\mathrm{Hz}$.

As in the 690- and 984-nm detection scheme, the 1088-nm scheme is also affected by population trapping in uncoupled magnetic sublevels when a fixed laser polarization is used. This problem can be overcome by periodically switching the laser polarization. A detailed description of the polarization-switching scheme is provided in Sec.~\ref{sec_2026_7_15_1}.

\section{\label{sec_2026_2_22_1}Results and Discussion}
\begin{table}[t]
\caption{Parameters of the CW 148-nm VUV laser adopted in the simulations of the direct 
			excitation of the $^{229}$Th nuclear isomer in Sec.~\ref{sec_2026_2_22_2}.}
	\label{tab_VUV_parameters}
	\begin{tabular}{lc}
		\hline
		\hline
		Parameter & Value \\
		\hline
		Laser power & $100~\mathrm{nW}$ \cite{Xiao2026} \\
		Focused beam area & $20~\mu\mathrm{m}^2$ \\
		Laser intensity & $0.5~\mathrm{W/cm^2}$ \\
		Laser linewidth & $\gamma_{n,l}/2\pi=10^{-3}$--10 kHz \\
		Nuclear Rabi frequency & $\Omega_n/2\pi=-3.4~\mathrm{Hz}$ \\
		Envelope function & $f_n(t)=1$ \\
		\hline
		\hline
	\end{tabular}
\end{table}

\subsection{\label{sec_2026_2_22_2} Direct VUV laser excitation of $^{229\mathrm{m}}\mathrm{Th}^{3+}$}
\begin{figure}[b]
	\includegraphics[width=3.5in]{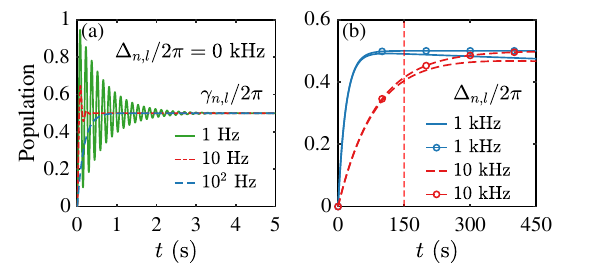}
	\caption{\label{Fig:2026_2_22_1} Isomeric population of the state $\ket{2}$ as a function of irradiation time (a) at zero detuning $\Delta_{n,l}/2\pi=0~\mathrm{kHz}$ for different laser linewidths $\gamma_{n,l}$, and (b) for different detunings $\Delta_{n,l}$ with the laser linewidth fixed at $\gamma_{n,l}/2\pi=10~\mathrm{kHz}$. In panel (b), the circled curves represent the analytic solutions in Eq.~\eqref{2025_12_8_4}, while the uncircled  curves correspond to the numerical solutions of Eq.~\eqref{2025_11_29_7}. The intensity of the 148-nm  laser is fixed at $0.5~\mathrm{W/cm^2}$ and the system is initially prepared in the state $\ket{1}$ for all cases.
	}
\end{figure}

\begin{figure}[t]
	\includegraphics[width=3.5in]{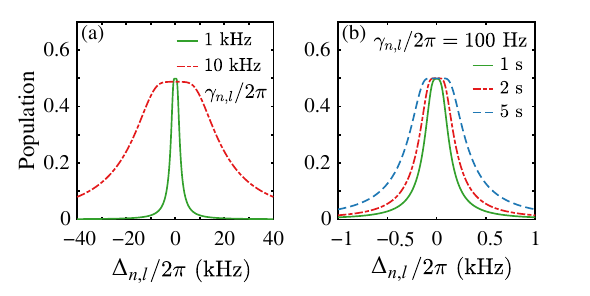}
	\caption{\label{Fig:2026_2_22_2}  Quasi-steady-state population of the isomeric state $\ket{2}$ as a function of the laser detuning $\Delta_{n,l}$ (a) for different laser linewidths, and (b) for different irradiation times, as labeled. In panel (a), the irradiation time is set to 20 and 200 s for $\gamma_{n,l}/2\pi=1$ and 10 kHz, respectively. In panel (b), the laser linewidth is fixed at $\gamma_{n,l}/2\pi=100~\mathrm{Hz}$.
		The intensity of the 148-nm laser is set to $0.5~\mathrm{W/cm^2}$ and the system is initially prepared in the state $\ket{1}$ for all cases.
	}
\end{figure}
We consider the direct excitation of the $^{229}$Th isomer using a CW 148-nm laser, for which the envelope function is constant [namely, $f_n(t)=1$]. With a laser power of 100~nW \cite{Xiao2026} focused onto an area of $20~\mu\mathrm{m}^2$, the laser intensity is calculated to be $0.5~\mathrm{W/cm^2}$, yielding a nuclear Rabi frequency of $\Omega_n/2\pi = -3.4~\mathrm{Hz}$. The corresponding VUV laser parameters are summarized in Table~\ref{tab_VUV_parameters}. These parameters are based on experimentally available CW VUV laser systems and are consistent with realistic  trapped-ion conditions.

By numerically solving Eq.~\eqref{2025_11_29_7}, we obtain the time-dependent population of the nuclear isomeric level for the zero-detuning case $\Delta_{n,l}/2\pi=0$ kHz, as shown in Fig.~\ref{Fig:2026_2_22_1}(a). When the laser linewidth is smaller than the Rabi frequency, clear Rabi oscillations are observed. As the linewidth increases, the oscillations gradually disappear. Due to the finite laser linewidth, the system eventually approaches the quasi-steady state with an isomeric population of $1/2$.
In Fig.~\ref{Fig:2026_2_22_1}(b), we compare the numerical solutions of Eq.~\eqref{2025_11_29_7} with the approximate analytic solutions in Eq.~\eqref{2025_12_8_4} for a fixed laser linewidth $\gamma_{n,l}/2\pi=10~\mathrm{kHz}$ but with different laser detunings $\Delta_{n,l}$. The two results agree well for times up to about $150~\mathrm{s}$, after which deviations gradually become apparent. This slight discrepancy arises because the numerical solution accounts for nuclear decay. 
For larger detuning, as described by the characteristic time $T_0$ in Eq.~\eqref{2025_12_8_5}, the time required to reach the quasi-steady state becomes longer.

The quasi-steady-state population of the isomeric state $\ket{2}$ is calculated as a function of the laser detuning $\Delta_{n,l}$ by numerically solving Eq.~\eqref{2025_11_29_7}, with the laser intensity fixed at $0.5~\mathrm{W/cm^2}$. 
In Fig.~\ref{Fig:2026_2_22_2}(a), the results for different laser linewidths are compared. According to the characteristic time $T_0$ [Eq.~\eqref{2025_12_8_5}], larger linewidths require longer irradiation times to reach the quasi-steady state. Accordingly, irradiation times of $20$ and $200~\mathrm{s}$ are chosen for $\gamma_{n,l}/2\pi = 1$ and $10~\mathrm{kHz}$, respectively, ensuring that the isomeric population exceeds $43\%$ at $\Delta_{n,l}=\gamma_{n,l}$ in both cases. Figure~\ref{Fig:2026_2_22_2}(b) shows that for a fixed linewidth of $100$~Hz, shorter irradiation times yield narrower excitation peaks. This behavior arises because, for shorter irradiation times, the quasi-steady state is achieved only at smaller detunings. For irradiation times of $1$, $2$, and $5~\mathrm{s}$, the isomeric population reaches $43\%$ at detunings $\Delta_{n,l}/2\pi$ of about $70$, $110$, and $180~\mathrm{Hz}$, respectively.

\subsection{\label{sec_2026_7_13_2}Detection of $^{229\mathrm{m}}\mathrm{Th}^{3+}$ via 690- and 984-nm fluorescence}

As illustrated in Fig.~\ref{HyperfineLevel}, the hyperfine splittings of the nuclear ground and isomeric levels differ by several hundred MHz. To ensure nuclear-level selectivity in the detection scheme, we assume that the electronic population transfer becomes negligible when the laser detuning exceeds 100 MHz. Under this condition, significant population transfer occurs only between electronic levels associated with the target nuclear level. Given the uncertainties in the hyperfine constants $A$ and $B$, the relevant hyperfine levels for the scheme using 690- and 984-nm lasers are uncertain by several MHz (see Appendix~\ref{app_hfs_uncertainty}). As summarized in Table~\ref{tab_2026_7_20_1}, the uncertainty does not exceed 5~MHz for the 690-nm transition and does not exceed approximately 10~MHz for the 984-nm transition. To assess the robustness of the scheme against these uncertainties, we therefore require the 690- and 984-nm lasers to maintain appreciable population transfer at detunings of 5~MHz and 10~MHz, respectively. 

We then investigate fluorescence-based population transfer between the hyperfine levels $\ket{a_m}$ and $\ket{c_m}$ via the intermediate state $\ket{b}$, where $m=0,\pm1$. As discussed in Sec.~\ref{sec_2026_7_13_1}, dark magnetic sublevels exist in this system under fixed-polarization laser driving. To prevent population trapping, we employ a polarization-switching scheme within a seven-level model that explicitly includes all relevant magnetic sublevels [see Eq.~\eqref{2026_3_20_1}]. The linearly, left-circularly, and right-circularly polarized components are applied sequentially, so that $\ket{a_m}$ or $\ket{c_m}$ sublevels are driven in turn. Experimentally, this scheme can be implemented by splitting a single laser beam into three paths with predefined polarizations and independently controlling their temporal sequences using acousto-optic modulators, such that only one polarization component is applied at a time. This time-ordered driving suppresses population trapping in dark states associated with any fixed polarization and avoids stable coherent population trapping under simultaneous coherent driving.

\begin{figure}[t]
	\includegraphics[width=3.5in]{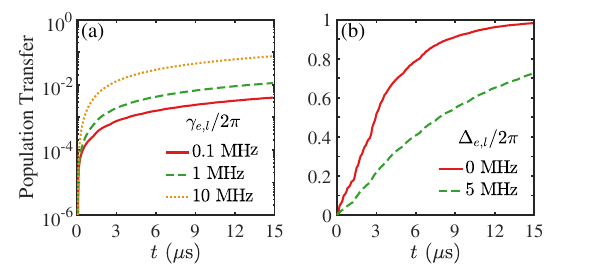}
	\caption{\label{Fig:2026_2_21_1} Total population transfer to the three magnetic sublevels $\ket{c_m}$ as a function of irradiation time (a) at a fixed detuning $\Delta_{e,l}/2\pi=100~\mathrm{MHz}$ for different laser linewidths $\gamma_{e,l}$, and (b) for different detunings $\Delta_{e,l}$ with the laser linewidth fixed at $\gamma_{e,l}/2\pi=0.1~\mathrm{MHz}$.  In all cases, the electronic population transfer is driven by a 690-nm laser with an intensity of $1~\mathrm{W/cm^2}$, and the laser polarization is switched every 1.25~$\mu$s. The system is initially uniformly distributed over the three magnetic sublevels $\ket{a_m}$.
	}
\end{figure}

For the 690-nm pumping step, a CW laser is temporally gated to generate a sequence of polarization-selective pulses. During the $i$th detection cycle, the linearly polarized component is applied over $t_i<t<t_i+1.25~\mu\mathrm{s}$, followed by the left-circularly polarized component over $t_i+1.25~\mu\mathrm{s}<t<t_i+2.5~\mu\mathrm{s}$ and the right-circularly polarized component over $t_i+2.5~\mu\mathrm{s}<t<t_i+3.75~\mu\mathrm{s}$. The polarization sequence is repeated four times, giving a total irradiation time of $15~\mu\mathrm{s}$. The envelope function is set to $f_e(t)=1$ while the corresponding polarization component is applied and to zero otherwise. The laser intensity is fixed at $1~\mathrm{W/cm^2}$, corresponding to a Rabi frequency of $\Omega_e^{(q)}/2\pi = -1.7~\mathrm{MHz}$. The population transfer from $\ket{a_m}$ to $\ket{c_m}$ is defined as 
\begin{equation}\label{2026_7_15_1}
P_c(t)=\sum_{m=-1}^{1}\rho_{c_m c_m}(t)~.
\end{equation}
By numerically solving Eq.~\eqref{2026_3_20_1} under these conditions, we obtain the time evolution of $P_c(t)$, as shown in Fig.~\ref{Fig:2026_2_21_1}. Figure~\ref{Fig:2026_2_21_1}(a) presents the results at a fixed laser detuning of $\Delta_{e,l}/2\pi=100~\mathrm{MHz}$ for different laser linewidths. A linewidth of 10 MHz produces a final population in $\ket{c_m}$ exceeding 5\% after 15~$\mu$s, whereas reducing the linewidth below 1 MHz suppresses the transferred population to less than 2\%. As shown in Fig.~\ref{Fig:2026_2_21_1}(b), for a fixed linewidth of 0.1~MHz, the transfer efficiency remains above 70\% within $15~\mu\mathrm{s}$, even at a detuning of 5 MHz. 

\begin{figure}[t]
	\includegraphics[width=3.5in]{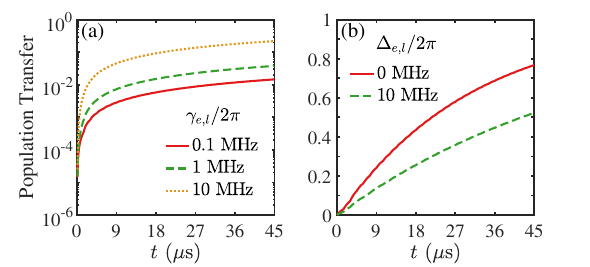}
	\caption{\label{Fig:2026_2_21_2}  Total population transfer to the three magnetic sublevels $\ket{a_m}$ as a function of irradiation time (a) at a fixed detuning $\Delta_{e,l}/2\pi=100~\mathrm{MHz}$ for different laser linewidths $\gamma_{e,l}$, and (b) for different detunings $\Delta_{e,l}$ with the laser linewidth fixed at $\gamma_{e,l}/2\pi=0.1~\mathrm{MHz}$.  In all cases, the electronic population transfer is driven by a 984-nm laser with an intensity of $0.4~\mathrm{W/cm^2}$, and the laser polarization is switched every 1.5~$\mu$s. The system is initially uniformly distributed over the three magnetic sublevels $\ket{c_m}$.
	}
\end{figure}

Following the 690-nm pumping step, a CW 984-nm laser is temporally gated to generate a sequence of polarization-selective pulses. During the $i$th detection cycle, the linearly polarized component is applied over $t_i+15~\mu\mathrm{s}<t<t_i+16.5~\mu\mathrm{s}$, followed by the left-circularly polarized component over $t_i+16.5~\mu\mathrm{s}<t<t_i+18~\mu\mathrm{s}$ and the right-circularly polarized component over $t_i+18~\mu\mathrm{s}<t<t_i+19.5~\mu\mathrm{s}$. This polarization sequence is repeated ten times, giving a total irradiation time of $45~\mu\mathrm{s}$. The envelope function is set to $f_e(t)=1$ while the corresponding polarization component is applied and to zero otherwise. The laser intensity is fixed at $0.4~\mathrm{W/cm^2}$, corresponding to a Rabi frequency of $\Omega_e^{(q)}/2\pi = -5.0~\mathrm{MHz}$. The population transfer from $\ket{c_m}$ to $\ket{a_m}$ is defined as
\begin{equation}\label{2026_7_15_2}
	P_a(t)=\sum_{m=-1}^{1}\rho_{a_m a_m}(t)~.
\end{equation}
Similarly, we numerically solve Eq.~\eqref{2026_3_20_1} under the above conditions to obtain $P_a(t)$, with the results shown in Fig.~\ref{Fig:2026_2_21_2}. Figure~\ref{Fig:2026_2_21_2}(a) shows the dependence of the population transferred to $\ket{a_m}$ on the 984-nm laser linewidth at a fixed detuning of $\Delta_{e,l}/2\pi=100~\mathrm{MHz}$. The results exhibit a trend similar to that found for the 690-nm pumping step.  As shown in Fig.~\ref{Fig:2026_2_21_2}(b), for a fixed laser linewidth of 0.1~MHz, a transfer efficiency exceeding 50\% is achieved within 45~$\mu$s, even in the presence of a 10-MHz detuning.

The above discussion shows that, with the laser parameters used in Figs.~\ref{Fig:2026_2_21_1}(b) and \ref{Fig:2026_2_21_2}(b), the detection scheme is robust against experimentally relevant uncertainties. We therefore adopt these parameters to estimate the fluorescence emission rates $R_{\mathrm{em}}$. Each detection cycle consists of a 690-nm pumping step followed by a 984-nm pumping step, with a total duration of $60~\mu\mathrm{s}$, corresponding to approximately $1.67\times10^4$ cycles per second.  After several cycles, the system reaches a stable periodic state. The numbers of 690- and 984-nm fluorescence photons emitted during one stable cycle are then given by
\begin{equation}\label{2026_7_15_3}
\begin{split}
	N_{690}
	= & 
\sum_{m=-1}^{1}\gamma_{e,ba_{m}}\int_{t_i}^{t_{i+1}}
		\rho_{bb}(t)\,\dx t ,\\
		N_{984}=	& 	\sum_{m=-1}^{1}\gamma_{e,bc_{m}}\int_{t_i}^{t_{i+1}}
		\rho_{bb}(t)\,\dx t .
	\end{split}
\end{equation}
For laser detunings of 5~MHz at 690 nm and 10~MHz at 984 nm, we obtain $N_{690}=0.50$ and $N_{984}=3.7$ photons per cycle, respectively. When both lasers are on resonance, the corresponding photon numbers are $N_{690}=0.86$ and $	N_{984}=6.5$. The larger 984-nm photon yield primarily arises from the stronger spontaneous-decay channel from $\ket{b}$ to  $\ket{c_m}$ . Accordingly, the on-resonance fluorescence emission rates are approximately $R_{\mathrm{em}}^{690}\approx1.4\times10^{4}~\mathrm{s}^{-1}$ per ion at 690 nm and $R_{\mathrm{em}}^{984}\approx1.1\times10^{5}~\mathrm{s}^{-1}$ per ion at 984 nm. For detunings of $5~\mathrm{MHz}$ at 690 nm and $10~\mathrm{MHz}$ at 984 nm, the corresponding emission rates remain approximately $R_{\mathrm{em}}^{690}\approx8.3\times10^{3}~\mathrm{s}^{-1}$ per ion and $R_{\mathrm{em}}^{984}\approx6.2\times10^{4}~\mathrm{s}^{-1}$ per ion, respectively. 

To estimate the experimentally observable fluorescence signal, we account for the finite efficiency of the fluorescence collection and imaging system. The detected fluorescence count rate, $R_\mathrm{det}$, is then given by
\begin{equation}\label{2026_7_19_1}
	R_\mathrm{det}=R_\mathrm{em}N_\mathrm{ion}\eta_\mathrm{det},
\end{equation}
where $N_\mathrm{ion}$ is the number of ions participating in the fluorescence cycle and \(\eta_\mathrm{det}\) is the overall detection efficiency. Following Ref.~\cite{Scharl2023}, we take \(N_\mathrm{ion}=50\), which is experimentally feasible. For the relevant fluorescence channels, we adopt overall detection efficiencies of \(\eta_\mathrm{det}^{690}=1.27\%\) and \(\eta_\mathrm{det}^{984}=0.35\%\) (see Appendix~\ref{app_2026_7_19_1}). Using the fluorescence emission rates obtained above, we find on-resonance detected count rates of \(R_\mathrm{det}^{690}\approx 8.9\times10^3~\mathrm{s^{-1}}\) and \(R_\mathrm{det}^{984}\approx 1.9\times10^4~\mathrm{s^{-1}}\) for 50 ions. At detunings of 5 MHz for the 690-nm laser and 10 MHz for the 984-nm laser, the corresponding rates remain \(R_\mathrm{det}^{690}\approx 5.3\times10^3~\mathrm{s^{-1}}\) and \(R_\mathrm{det}^{984}\approx 1.1\times10^4~\mathrm{s^{-1}}\). 

\subsection{\label{sec_2026_7_15_1}Detection of $^{229\mathrm{m}}\mathrm{Th}^{3+}$ via 1088-nm fluorescence}
\begin{figure}[t]
	\includegraphics[width=3.5in]{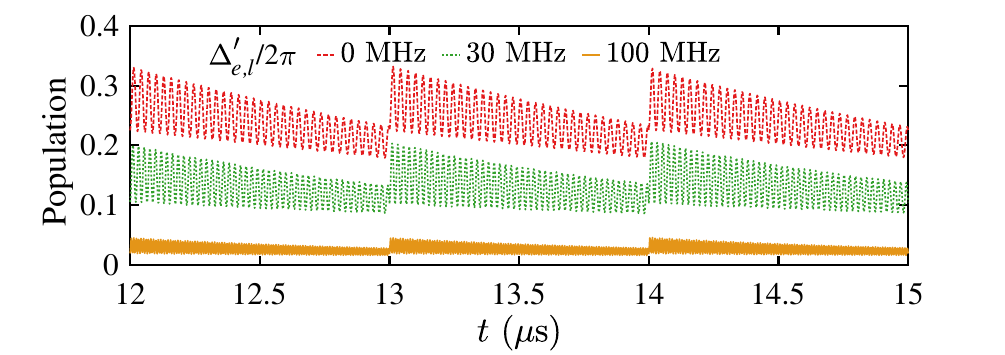}
	\caption{\label{Fig:2026_2_21_3}  Electronic population of the state $\ket{b^\prime}$ as a function of the irradiation time for different 1088-nm laser detunings $\Delta_{e,l}^\prime/2\pi$, as labeled. The laser linewidth is fixed at $\gamma_{e,l}^\prime/2\pi=0.1~\mathrm{MHz}$, and the laser intensity is set to $5~\mathrm{W/cm^2}$. Initially, the population is uniformly distributed among the three magnetic sublevels $\ket{a^\prime_m}$. The laser polarization is switched every 1~$\mu$s. After the first few cycles, the system evolves into a stable periodic solution. Therefore, the evolution time is chosen to be between 12 and 15~$\mu$s. 
	}
\end{figure}
We now investigate the population transfer from $\ket{a^\prime_m}$ to $\ket{b^\prime}$ using Eq.~\eqref{2026_7_22_3}. A CW 1088-nm laser with an intensity of $5~\mathrm{W/cm^2}$ is applied at several detunings
$\Delta_{e,l}^\prime/2\pi$. Our calculations for the 690- and 984-nm scheme show that a linewidth of 0.1~MHz is sufficiently narrow for hyperfine-resolved excitation. We therefore adopt the same value, $\gamma_{e,l}^\prime/2\pi=0.1~\mathrm{MHz}$, for the 1088-nm scheme. As in the 690- and 984-nm
detection scheme, the CW laser is gated by acousto-optic modulators to
generate a sequence of polarization-selective pulses, thereby preventing
population trapping in uncoupled magnetic dark states. The laser polarization
is switched every $1~\mu\mathrm{s}$.  The envelope function is set to $f_e(t)=1$ while the corresponding polarization component is applied and to zero otherwise. The associated Rabi frequency is $\Omega_e ^{(q)}/2\pi= -17.3~\mathrm{MHz}$.  After several polarization-switching cycles, the system evolves into a stable periodic state. We therefore focus on the interval from $12$ to
$15~\mu\mathrm{s}$, which corresponds to one complete period of the stable
periodic evolution. The resulting population of the state $\ket{b^\prime}$ as
a function of irradiation time for different laser detunings is shown in
Fig.~\ref{Fig:2026_2_21_3}.

As summarized in Table~\ref{tab_2026_7_20_1} of Appendix~\ref{app_hfs_uncertainty}, the relevant 1088-nm transition
frequencies differ by several hundred MHz among different nuclear-state configurations and hyperfine channels, whereas the remaining experimental uncertainty is approximately $30~\mathrm{MHz}$. We therefore show in Fig.~\ref{Fig:2026_2_21_3} the population dynamics for laser detunings of $\Delta_{e,l}^\prime/2\pi=0$, $30$, and $100~\mathrm{MHz}$. Whenever a given polarization component is applied, the system undergoes damped Rabi oscillations. For detunings of $0$, $30$, and $100~\mathrm{MHz}$, the peak populations of the state $\ket{b^\prime}$ are approximately $33.2\%$, $20.7\%$, and $4.3\%$, respectively. These results show that the chosen laser
parameters provide a favorable compromise between nuclear-state selectivity
and robustness against the current experimental frequency uncertainty.

The duration of one detection cycle in the 1088-nm detection scheme is
	$3~\mu\mathrm{s}$. Once the system reaches a stable periodic
	state, the number of 1088-nm fluorescence photons emitted during one
	complete cycle is given by
	\begin{equation}\label{2026_7_22_4}
		N_{1088}= \sum_{m=-1}^{1}\gamma_{e,b^\prime a^\prime_m} \int_{t_i}^{t_{i+1}}	\rho_{b^\prime b^\prime}(t)\dx t.
	\end{equation}
For laser detunings of $\Delta_{e,l}^{\prime}/2\pi=0$ and $30~\mathrm{MHz}$, the numbers of emitted photons per detection cycle are $ N_{1088}=0.66$ and $0.36$, respectively. The resulting fluorescence emission rates are therefore $R_{\mathrm{em}}^{1088}=2.2\times10^5~\mathrm{s}^{-1}$ per ion and $1.2\times10^5~\mathrm{s}^{-1}$ per ion  for detunings of $0$ and $30~\mathrm{MHz}$, respectively.

Using Eq.~\eqref{2026_7_19_1}, together with $N_{\mathrm{ion}}=50$ and a total detection efficiency of
$\eta_{\mathrm{det}}^{1088}=0.053\%$ [see Eq.~\eqref{2026_7_19_4}], we obtain detected fluorescence count rates of $R_{\mathrm{det}}^{1088}=5.8\times10^{3}~\mathrm{s}^{-1}$ and 	$3.2\times10^{3}~\mathrm{s}^{-1}$ for laser detunings of
$\Delta_{e,l}^{\prime}/2\pi=0$ and $30~\mathrm{MHz}$, respectively.

We also note that trap-induced frequency shifts \cite{Campbell2011ThoriumIV}, including contributions from magnetic-field fluctuations, trap electric fields, and excess micromotion, may perturb the hyperfine transition frequencies. Nevertheless, these shifts are expected to remain below 1 MHz, which is significantly smaller than the several-MHz detuning tolerance demonstrated in our simulations. Therefore, the proposed detection schemes are expected to be robust against these perturbations.

\subsection{Total search time for the isomeric energy}\label{SecTime}

To search for the isomeric transition energy, we assume a fixed irradiation time $T_{\mathrm{irr}}$ and a constant energy scan step $\Delta$. With a detection interval $T_{\mathrm{det}}$ following each irradiation period, the total search time $T_{\mathrm{tot}}$ is proportional to:
\begin{equation}\label{2026_3_30_1}
	T_{\mathrm{tot}}\propto \dfrac{T_{\mathrm{irr}}+ T_{\mathrm{det}}}{\Delta}.
\end{equation}
We assume the excitation probability of the isomer to reach a threshold value $P_{\mathrm{cri}}$ at time $T_{\mathrm{irr}}$, when the laser detuning is as large as the scan step $\Delta$. Based on Eq.~\eqref{2025_12_8_4}, the relationship between the irradiation time $T_{\mathrm{irr}}$ 
and the energy step $\Delta$ is given by
     \begin{equation}\label{2026_4_1_1}
T_{\mathrm{irr}} = -\frac{4\Delta^2 + \gamma_{n,l}^2}{8\gamma_{n,l}\Omega_n^2} \ln(1 - 2P_{\mathrm{cri}}).
\end{equation} 

According to Eq.~\eqref{2026_4_1_1}, $T_{\mathrm{tot}}$ is a function of the scan step $\Delta$. Minimizing the total search time via the condition $\partial T_{\mathrm{tot}}/\partial \Delta = 0$ yields the optimal energy scan step $\Delta_{\mathrm{opt}}$:
\begin{equation}\label{2026_4_1_3}
	\Delta_{\mathrm{opt}} = \sqrt{\frac{\gamma_{n,l}^2}{4} - \frac{2\gamma_{n,l}\Omega_n^2 T_{\mathrm{det}}}{\ln(1 - 2P_{\mathrm{cri}})}}.
\end{equation}
By substituting $\Delta_{\mathrm{opt}}$ into Eq.~\eqref{2026_4_1_1}, the optimal irradiation time $T_{\mathrm{irr}}^{\mathrm{opt}}$ is uniquely determined. 
The critical isomeric population $P_{\mathrm{cri}}$ is determined from a statistical analysis of the finite number of detected fluorescence photons, $N_{\mathrm{ph}}$, and therefore depends on the fluorescence-readout scheme. For the 690-nm fluorescence-detection method, we choose a detection time of $T_{\mathrm{det}} = 60$ ms, whereas $T_{\mathrm{det}} = 30$ ms is used for both the 984- and 1088-nm methods.
Based on our photon-counting statistical analysis, the corresponding critical isomeric populations are $P_{\mathrm{cri}}^{690}=45\%$, $P_{\mathrm{cri}}^{984}=44\%$, and $P_{\mathrm{cri}}^{1088}=42\%$. The detailed statistical calculations are presented in Appendix~\ref{app}.

Assuming a nuclear Rabi frequency of $\Omega_n/2\pi = -3.4~\text{Hz}$, we calculate the optimal energy scan steps $\Delta_{\mathrm{opt}}/2\pi$ using Eq.~\eqref{2026_4_1_3}. For the 690-nm  detection method, the optimal scan steps are 53.7, 503.8, and $5003.8~\mathrm{Hz}$ for $\gamma_{n,l}/2\pi=0.1$, 1, and $10~\mathrm{kHz}$, respectively. The corresponding optimal irradiation times $T_{\mathrm{irr}}^{\mathrm{opt}}$ are 0.85, 7.99, and $79.3~\mathrm{s}$, respectively. For an isomeric energy uncertainty of $100~\text{MHz}$, the resulting total search times $T_{\mathrm{tot}}$ are approximately 19.7, 18.5, and 18.4 days, respectively.

For the 984-nm detection method, the optimal scan steps are 52.0, 502.1, and $5002.1~\mathrm{Hz}$ for $\gamma_{n,l}/2\pi=0.1$, 1, and $10~\mathrm{kHz}$, respectively. The corresponding optimal irradiation times $T_{\mathrm{irr}}^{\mathrm{opt}}$ are 0.76, 7.33, and $73.0~\mathrm{s}$, respectively. For an isomeric transition-frequency uncertainty of $100~\mathrm{MHz}$, the resulting total search times $T_{\mathrm{tot}}$ are approximately 17.6, 17.0, and 16.9 days, respectively.

For the 1088-nm detection method, the optimal scan steps are 52.3, 502.4, and $5002.4~\mathrm{Hz}$ for $\gamma_{n,l}/2\pi=0.1$, 1, and $10~\mathrm{kHz}$, respectively. The corresponding optimal irradiation times $T_{\mathrm{irr}}^{\mathrm{opt}}$ are 0.66, 6.34, and $63.1~\mathrm{s}$, respectively. For an isomeric transition-frequency uncertainty of $100~\mathrm{MHz}$, the resulting total search times $T_{\mathrm{tot}}$ are approximately 15.3, 14.7, and 14.6 days, respectively. In the above estimates, the system-reset time following each scan step is estimated to be approximately $1~\mathrm{ms}$ (see Appendix~\ref{app_initial_state}) and is therefore neglected.

\section{\label{sec_2020_10_12_3}Conclusion}

In summary, we develop a quantum-optical framework for hyperfine-resolved laser excitation and detection of the $^{229}$Th nuclear isomer in trapped $^{229}$Th$^{3+}$ ions. Within this framework, we analyze the roles of laser linewidth, detuning, and irradiation time, and show that their proper matching is important for efficient isomeric excitation. We investigate three detection methods based on electronic fluorescence at 690, 984, and 1088 nm. The 690-nm and 984-nm scheme enables indirect detection of the nuclear state through fluorescence suppression in the nuclear-ground-state configuration,  whereas the 1088-nm scheme directly probes the isomeric configuration.
For 50 trapped ions, the predicted detectable photon count rates are on the order of $10^{3}~\mathrm{s}^{-1}$ at 690 nm, $10^{4}~\mathrm{s}^{-1}$ at 984 nm, and  $10^{3}~\mathrm{s}^{-1}$ at 1088 nm.
Based on the quasi-steady-state solution, we identify a trade-off between the irradiation time per frequency step and the scan-step size in resonance searches. For an isomeric energy uncertainty of 100 MHz, the nuclear transition can be located within about one month using currently available CW 148-nm VUV laser parameters. Trapped $^{229}$Th$^{3+}$ ions are promising systems for nuclear precision metrology, which requires reliable preparation, excitation, and detection of the nuclear state. Our results therefore provide practical guidance for near-future trapped-ion spectroscopy of the $^{229}$Th nuclear transition and support the development of trapped-ion nuclear clocks.

\section*{ACKNOWLEDGMENTS}
Wu Wang acknowledges helpful
discussions with Prof. Yong Li and Assoc. Prof. Fen Zou. This work is supported by the Strategic Priority Research Program of the Chinese Academy of Sciences (Grant No. XDB0920000).

\section*{DATA AVAILABILITY}
The data supporting this study's findings are available
within the article.

\appendix

\section{Preparation of the initial hyperfine state}
\label{app_initial_state}

The initial state \((I_g,5f_{5/2},F=1)\) can be prepared by optical pumping through the 1088-nm transitions between the
\((I_g,5f_{5/2},F)\) and \((I_g,6d_{3/2},F)\) hyperfine manifolds. Specifically, the transitions $(I_g,5f_{5/2},F=a)
\rightarrow (I_g,6d_{3/2},F=a-1)$ are sequentially driven for \(a=5,4,\) and \(3\). As in the detection schemes described in Sec.~\ref{sec_2026_7_13_1}, population trapping in magnetic dark states is avoided by periodically switching the laser polarization among linear, left-circular, and right-circular polarizations, with each polarization component applied for \(1~\mu\mathrm{s}\). A Rabi frequency of \(5~\mathrm{MHz}\) is assumed. After these three pumping steps, the remaining population is confined to the
\((I_g,5f_{5/2},F=0,1,2)\) hyperfine levels. The transitions $(I_g,5f_{5/2},F=2)\rightarrow(I_g,6d_{3/2},F=1)$ and $(I_g,5f_{5/2},F=0)\rightarrow(I_g,6d_{3/2},F=1)$ are then driven alternately. Repeating this two-step pumping cycle five times transfers the remaining population into
\((I_g,5f_{5/2},F=1)\).

The required irradiation time for each pumping transition is determined by the decay rates from the corresponding excited hyperfine level to lower hyperfine states other than the initially addressed state. Specifically, for $(I_g,5f_{5/2},F=a)\rightarrow(I_g,6d_{3/2},F=a-1)$ with $a = 5,4,3$, and 2, the pumping efficiency depends on the decay rates from \((I_g,6d_{3/2},F=a-1)\) to lower hyperfine levels other than
\((I_g,5f_{5/2},F=a)\). Similarly, for \((I_g,5f_{5/2},F=0)\rightarrow(I_g,6d_{3/2},F=1)\), the required irradiation time is determined by the decay rates from
\((I_g,6d_{3/2},F=1)\) to lower hyperfine levels other than
\((I_g,5f_{5/2},F=0)\). Therefore, the required pumping duration depends on the specific hyperfine transition.

Our estimates show that population-transfer efficiencies above \(99\%\) can be achieved with irradiation times of approximately \(180\), \(100\), \(70\), and \(50~\mu\mathrm{s}\) for the transitions $(I_g,5f_{5/2},F=a)\rightarrow(I_g,6d_{3/2},F=a-1)$ with \(a=5,4,3,\) and \(2\), respectively. The pumping step $(I_g,5f_{5/2},F=0)\rightarrow(I_g,6d_{3/2},F=1)$ requires approximately \(50~\mu\mathrm{s}\). Therefore, the total preparation time is estimated as
$T_{\rm prep}=180+100+70+5\times(50+50)=850~\mu\mathrm{s}.$

After a detection cycle, the initial state can be restored using a similar optical-pumping procedure. 

\section{\label{app_2026_3_18_1}Derivation of Eq.~(\ref{2025_12_6_1})}
The nucleus-laser interaction can be expanded as \cite{Wang2023}
\begin{equation}\label{2026_3_19_1}
	H_{nl} = -\sqrt{4\pi} \sum_{\tau L} \sqrt{[L]} \mathcal{M}^{(\tau L)}\cdot C^{(\tau L)},
\end{equation}
where $\mathcal{M}^{(\tau L)}$ denotes the nuclear multipole transition operator of rank $L$ \cite{Schwartz1955}, and $C^{(\tau L)}$ is the corresponding expansion coefficient, whose general component $C^{(\tau L)}_m$ is given in Ref.~\cite{Wang2023}. Here, Coulomb gauge is used and $\tau = E, M$ specifies whether the transition is electric or magnetic. 

For the nuclear $M1$ transition, $C^{(M1)}_m$ is given by
\begin{equation}\label{2026_3_19_2}
	C^{(M1)}_m = -\dfrac{ik\sqrt{2}}{3}(w(t)i+\mathrm{c.c}.)\dfrac{\hat{\bm e}_z}{2}\cdot\bm A^{\mathrm{M}}_{1m}(\hat{{\bm k}}),
\end{equation}
where $w(t)$ is a temporal envelope function associated with the vector potential $\bm A(t)$ (see Ref.~\cite{Wang2023}), and $\bm A^{\mathrm{M}}_{1m}(\hat{{\bm k}})=\bm Y_{11m}(\hat{{\bm k}})$ is a vector spherical harmonic \cite{R.Johnson2007}. Here, $k$ is the wave number and $\hat{\bm e}_z$ is the unit vector along the $z$ axis. Assuming that the laser electric field takes the form $\bm E(t)=\hat{\bm e}_z F_{n,0} f_n(t)\cos(\omega_l t)$, the corresponding vector potential is written as
\begin{equation}\label{2026_3_19_3}
	\bm A(t)=\hat{\bm e}_z\dfrac{ F_{n,0}}{2}\bigg(\dfrac{f_n(t)e^{-i\omega_l t}}{ik}+\mathrm{c.c.}\bigg).
\end{equation}
Accordingly, one obtains $w(t) = f_n(t)e^{-i\omega_l t}/(i k)$.
It follows from the definition of $\bm Y_{11m}(\hat{{\bm k}})$ that
\begin{equation}\label{2026_3_19_4}
	\hat{\bm e}_z\cdot\bm A^{\mathrm{M}}_{1m}(\hat{{\bm k}})=
\begin{cases}
	-\frac{1}{4} i \sqrt{\frac{3}{\pi }}, &  m=1,\\
	0, &  m=0,\\
		\frac{1}{4} i \sqrt{\frac{3}{\pi }},&m=-1,
\end{cases}
\end{equation}
where $\hat{\bm k}$ is taken along the $y$ axis.

By using Eqs.~\eqref{2026_3_19_1}, \eqref{2026_3_19_2}, and \eqref{2026_3_19_4}, the nucleus-laser interaction for the $M1$ transition is expressed as
\begin{equation}\label{2026_3_19_5}
	H_{nl}^{(M1)}=\dfrac{1}{\sqrt{2}}\left(\mathcal{M}^{(M1)}_1	-\mathcal{M}^{(M1)}_{-1}\right) f_n(t)F_{n,0}\cos(\omega_l t).
\end{equation}
The  Rabi frequency in the Hamiltonian~\eqref{2025_12_4_1} is calculated by
\begin{equation}\label{2026_3_19_6}
	\Omega_n(t)=\melb{I_e5f_{5/2};F=1m_2}{H_{nl}^{(M1)}}{I_g5f_{5/2};F=1m_1}.
\end{equation}
By applying the Wigner-Eckart theorem and RWA, Eq.~\eqref{2025_12_6_1} is obtained from Eq.~\eqref{2026_3_19_6}, where $M_n$ is calculated to be
\begin{equation}\label{2025_12_5_1}
	M_n=3 (-1)^{I_e+5/2}  \left\{\begin{matrix}
	I_e& I_g& 1 \\
	1 & 1 & 5/2
\end{matrix}\right\}\melb{I_e }{\big|\mathcal{M}^{(M1)}\big|}{I_g }.
\end{equation}
Here, the reduced matrix element $\melb{I_e }{\big|\mathcal{M}^{(M1)}\big|}{I_g }$ is related to $B(M1,I_e\rightarrow I_g)$ via
\begin{equation}\label{2026_3_19_8}
B(M1,I_e\rightarrow I_g)=\dfrac{3}{4\pi}\dfrac{\big|\melb{I_e }{\big|\mathcal{M}^{(M1)}\big|}{I_g }\big|^2}{2I_e+1}.
\end{equation}

\section{\label{app_2026_3_22_1}Expression of $M_e$}
The explicit expression for $M_e$ in Eq.~\eqref{2025_12_6_4} is given by
\begin{equation}\label{2026_3_22_1}
	\begin{split}
			M_e = & (-1)^{I + 1 + J_i + F_f} \sqrt{[F_i][F_f]} \left\{ \begin{matrix}
				J_f & J_i & 1 \\
				F_i & F_f & I
			\end{matrix} \right\} \times\\
		&\melb{\Gamma_{f}J_{f}}{\big|\mathcal{O}^{(E1)}\big|}{\Gamma_{i}J_{i}}.
	\end{split}
\end{equation}
The decay rate $\gamma$ between the electronic levels $(I,\Gamma_i J_i)$ and $(I,\Gamma_f J_f)$ (without resolving the hyperfine structure) is related to the decay rate $\gamma_e$ in Eq.~\eqref{2025_12_6_6} by (see, e.g., Ref.~\cite{Wang2024})
\begin{equation}\label{2025_12_6_7}
	\gamma_e= \gamma(2F_f+1)(2J_i+1)  \left\{\begin{matrix}
		J_f& J_i & 1 \\
		F_i & F_f & I
	\end{matrix}\right\}^2.
\end{equation}
The decay rate $\gamma$ is determined by the corresponding electronic lifetime $\tau$ via $\gamma=1/\tau$. Therefore, once $\tau$ is known, both $\gamma_e$ and $M_e$ are fully determined.
\section{\label{app_hfs_uncertainty}Uncertainty in hyperfine-resolved transition frequencies}

\begin{table*}[!htbp]
    \centering
\renewcommand{\arraystretch}{1.15}
\setlength{\tabcolsep}{8pt}
\caption{Hyperfine frequency offsets \(\delta\nu_\mathrm{hfs}\) and their propagated uncertainties for the relevant transitions in \(^{229}\mathrm{Th}^{3+}\) and \(^{229\mathrm{m}}\mathrm{Th}^{3+}\). Each entry is given as \(\delta\nu_\mathrm{hfs}\pm\sigma_\mathrm{hfs}\) in MHz, where \(\sigma_\mathrm{hfs}\) denotes the uncertainty in the transition frequency. The corresponding transition frequency is \(\nu=\nu_0+\delta\nu_\mathrm{hfs}\), where \(\nu_0\) is the associated fine-structure transition frequency.}
\label{tab_2026_7_20_1}
\begin{tabular}{p{0.20\textwidth}cccc}
	\hline
	\hline
	\noalign{\vskip 2pt}
	Channel & \multicolumn{2}{c}{\(^{229}\mathrm{Th}^{3+}\), \(I_g=5/2\)} & \multicolumn{2}{c}{\(^{229\mathrm{m}}\mathrm{Th}^{3+}\), \(I_e=3/2\)} \\[2pt]
	\cline{2-5}
	\noalign{\vskip 2pt}
	& Component (\(F_i\rightarrow F_f\)) & \(\delta\nu_\mathrm{hfs}\pm\sigma_\mathrm{hfs}\) & Component (\(F_i\rightarrow F_f\)) & \(\delta\nu_\mathrm{hfs}\pm\sigma_\mathrm{hfs}\) \\[2pt]
	\hline
	\noalign{\vskip 1pt}
	\multicolumn{5}{l}{690 nm: \(5f_{5/2}\rightarrow 6d_{5/2}\)} \\[2pt]
	& \(0\rightarrow 1\) & \(468.3\pm3.3\)    & \(1\rightarrow 1\) & \(-690.6\pm32.3\) \\
	& \(1\rightarrow 0\) & \(1591.0\pm3.4\)   & \(1\rightarrow 2\) & \(-1873.2\pm24.5\) \\
	& \(1\rightarrow 1\) & \(931.2\pm3.0\)    & \(2\rightarrow 1\) & \(626.0\pm23.9\) \\
	& \(1\rightarrow 2\) & \(-146.0\pm2.5\)   & \(2\rightarrow 2\) & \(-556.5\pm11.3\) \\
	& \(2\rightarrow 1\) & \(1652.6\pm2.7\)   & \(2\rightarrow 3\) & \(-1180.0\pm19.7\) \\
	& \(2\rightarrow 2\) & \(575.5\pm2.1\)    & \(3\rightarrow 2\) & \(449.4\pm14.3\) \\
	& \(2\rightarrow 3\) & \(-433.7\pm1.5\)   & \(3\rightarrow 3\) & \(-174.1\pm21.5\) \\
	& \(3\rightarrow 2\) & \(1146.8\pm1.9\)   & \(3\rightarrow 4\) & \(1142.3\pm15.9\) \\
	& \(3\rightarrow 3\) & \(137.6\pm1.2\)    & \(4\rightarrow 3\) & \(-641.7\pm21.3\) \\
	& \(3\rightarrow 4\) & \(-75.7\pm1.1\)    & \(4\rightarrow 4\) & \(674.8\pm15.7\) \\
	& \(4\rightarrow 3\) & \(-54.2\pm1.2\)    & -- & -- \\
	& \(4\rightarrow 4\) & \(-267.5\pm1.0\)   & -- & -- \\
	& \(4\rightarrow 5\) & \(1285.4\pm2.0\)   & -- & -- \\
	& \(5\rightarrow 4\) & \(-2039.7\pm1.5\)  & -- & -- \\
	& \(5\rightarrow 5\) & \(-486.8\pm2.3\)   & -- & -- \\
	\hline
	\multicolumn{5}{l}{984 nm: \(5f_{7/2}\rightarrow 6d_{5/2}\)} \\[2pt]
	& \(1\rightarrow 0\) & \(987.0\pm10.6\)   & \(2\rightarrow 1\) & \(-186.4\pm28.8\) \\
	& \(1\rightarrow 1\) & \(327.2\pm10.5\)   & \(2\rightarrow 2\) & \(-1369.0\pm19.7\) \\
	& \(1\rightarrow 2\) & \(-749.9\pm10.3\)  & \(2\rightarrow 3\) & \(-1992.4\pm25.4\) \\
	& \(2\rightarrow 1\) & \(992.9\pm7.6\)    & \(3\rightarrow 2\) & \(-170.0\pm9.0\) \\
	& \(2\rightarrow 2\) & \(-84.2\pm7.3\)    & \(3\rightarrow 3\) & \(-793.4\pm18.4\) \\
	& \(2\rightarrow 3\) & \(-1093.3\pm7.2\)  & \(3\rightarrow 4\) & \(523.0\pm11.4\) \\
	& \(3\rightarrow 2\) & \(641.3\pm4.8\)    & \(4\rightarrow 3\) & \(-162.0\pm21.3\) \\
	& \(3\rightarrow 3\) & \(-367.9\pm4.6\)   & \(4\rightarrow 4\) & \(1154.4\pm15.6\) \\
	& \(3\rightarrow 4\) & \(-581.2\pm4.5\)   & \(5\rightarrow 4\) & \(388.9\pm14.0\) \\
	& \(4\rightarrow 3\) & \(89.4\pm4.1\)     & -- & -- \\
	& \(4\rightarrow 4\) & \(-124.0\pm4.0\)   & -- & -- \\
	& \(4\rightarrow 5\) & \(1428.9\pm4.4\)   & -- & -- \\
	& \(5\rightarrow 4\) & \(-372.0\pm3.8\)   & -- & -- \\
	& \(5\rightarrow 5\) & \(1180.9\pm4.2\)   & -- & -- \\
	& \(6\rightarrow 5\) & \(-319.0\pm7.1\)   & -- & -- \\
	\hline
	\multicolumn{5}{l}{1088 nm: \(5f_{5/2}\rightarrow 6d_{3/2}\)} \\[2pt]
	& \(0\rightarrow 1\) & \(-101.5\pm9.2\)   & \(1\rightarrow 0\) & \(964.0\pm29.5\) \\
	& \(1\rightarrow 1\) & \(361.3\pm9.1\)    & \(1\rightarrow 1\) & \(-591.0\pm25.7\) \\
	& \(1\rightarrow 2\) & \(-1140.1\pm4.4\)  & \(1\rightarrow 2\) & \(-2413.0\pm25.5\) \\
	& \(2\rightarrow 1\) & \(1082.8\pm9.0\)   & \(2\rightarrow 1\) & \(725.6\pm13.7\) \\
	& \(2\rightarrow 2\) & \(-418.6\pm4.2\)   & \(2\rightarrow 2\) & \(-1096.4\pm13.3\) \\
	& \(2\rightarrow 3\) & \(-972.0\pm5.1\)   & \(2\rightarrow 3\) & \(-609.4\pm12.9\) \\
	& \(3\rightarrow 2\) & \(152.7\pm4.1\)    & \(3\rightarrow 2\) & \(-90.5\pm15.8\) \\
	& \(3\rightarrow 3\) & \(-400.7\pm5.0\)   & \(3\rightarrow 3\) & \(396.5\pm15.5\) \\
	& \(3\rightarrow 4\) & \(2032.5\pm5.1\)   & \(4\rightarrow 3\) & \(-71.1\pm15.3\) \\
	& \(4\rightarrow 3\) & \(-592.5\pm5.0\)   & -- & -- \\
	& \(4\rightarrow 4\) & \(1840.7\pm5.1\)   & -- & -- \\
	& \(5\rightarrow 4\) & \(68.5\pm5.2\)     & -- & -- \\
	\hline
	\hline
\end{tabular}
\end{table*}

The uncertainties in the hyperfine-resolved transition frequencies are obtained by propagating the uncertainties in the hyperfine constants \(A\) and \(B\) listed in Table~\ref{tab_2026_2_25_1}. The quoted uncertainties in \(A\) and \(B\) originate either from direct experimental measurements or from the nuclear magnetic-dipole and electric-quadrupole moments used to infer these constants. 

Using the hyperfine shift formula in Eq.~\eqref{2026_2_23_1}, the shift of each hyperfine level can be written as
\begin{equation}
	\Delta\nu_\mathrm{hfs}(F,A, B)=	E_{M1}(F,A)+E_{E2}(F,B).
\end{equation}
The frequency of a hyperfine-resolved transition can then be written explicitly in terms of the initial- and final-state hyperfine quantum numbers and constants as
\begin{equation}
	\nu(F_i,A_i,B_i,F_f,A_f,B_f)
	=\nu_0+\delta\nu_\mathrm{hfs},
\end{equation}
where
\begin{equation}
	\delta\nu_\mathrm{hfs}
	=\Delta\nu_{\mathrm{hfs}}(F_f,A_f,B_f)
	-\Delta\nu_{\mathrm{hfs}}(F_i,A_i,B_i)
\end{equation}
and \(\nu_0\) is the corresponding fine-structure transition frequency. The propagated uncertainty of transition frequency is therefore
\begin{equation}
	\begin{aligned}
	\sigma_\mathrm{hfs}
		&=
		\Bigg[
		\left(\frac{\partial\nu}{\partial A_i}\sigma_{A_i}\right)^2
		+\left(\frac{\partial\nu}{\partial B_i}\sigma_{B_i}\right)^2
		\\
		&\quad
		+\left(\frac{\partial\nu}{\partial A_f}\sigma_{A_f}\right)^2
		+\left(\frac{\partial\nu}{\partial B_f}\sigma_{B_f}\right)^2
		\Bigg]^{1/2}.
	\end{aligned}
\end{equation}
Here, \(\sigma_{A_i}\), \(\sigma_{B_i}\), \(\sigma_{A_f}\), and \(\sigma_{B_f}\) are the uncertainties of the corresponding hyperfine constants. The resulting \(\sigma_\mathrm{hfs}\) is the transition-frequency uncertainty arising from the uncertainties in the initial- and final-state hyperfine splittings.

Using the hyperfine constants listed in Table~\ref{tab_2026_2_25_1},  the predicted frequency offsets and corresponding propagated uncertainties for the 690-, 984-, and 1088-nm transitions are summarized in Table~\ref{tab_2026_7_20_1}. The table includes transitions associated with both the nuclear ground state, \(I_g=5/2\), and the isomeric state, \(I_e=3/2\). Only electric-dipole-allowed components with \(\Delta F=0,\pm1\) are listed.

\section{\label{app_2026_7_19_1}Detection efficiency \(\eta_\mathrm{det}\) }
The overall detection efficiency \(\eta_\mathrm{det}\) is wavelength dependent and accounts for the fluorescence collection efficiency determined by the collection solid angle, the optical transmission, the filter transmission, and the detector quantum efficiency. For the 690-nm fluorescence channel, we adopt the overall detection efficiency estimated in Ref.~\cite{Scharl2023}, $\eta_{\mathrm{det}}^{690}=1.27\%$. This value already accounts for the finite collection solid angle, transmission and reflection losses of the optical components, spectral filtering, and the quantum efficiency of the camera.

For the 984-nm channel, we use a fluorescence collection efficiency of approximately \(7\%\) \cite{Campbell2011ThoriumIV}, a silicon-camera quantum efficiency of \(20\%\), an additional optical throughput of \(50\%\), and a bandpass-filter transmission of \(50\%\). This gives
\begin{equation}\label{2026_7_19_3}
\eta_\mathrm{det}^{984}=0.07\times0.20\times0.50\times0.50=0.35\%.
\end{equation}

For the 1088-nm channel, detector selection is more demanding because silicon-based detectors have quantum efficiencies below \(1\%\) near this wavelength \cite{Campbell2011ThoriumIV}. As a conservative estimate for near-infrared single-photon detection, we consider an InGaAs/InP avalanche photodiode with a quantum efficiency of \(3\%\). Assuming the same fluorescence collection efficiency, optical throughput, and filter transmission as for the 984-nm channel, the resulting detection efficiency is
\begin{equation}\label{2026_7_19_4}
	\eta_\mathrm{det}^{1088}=0.07\times0.03\times0.50\times0.50=0.053\%.
\end{equation}

\section{Statistical model for fluorescence readout}
\label{app}

For an isomeric excitation probability \(P\), the number \(K\) of ions in the nuclear isomeric state is assumed to follow a binomial distribution:
\begin{equation}\label{2026_8_1_1}
\mathrm{Pr}(K=k)
=
\binom{N_\mathrm{ion}}{k}P^k(1-P)^{N_\mathrm{ion}-k},
\end{equation}
where $k$ is an integer satisfying $0\leqslant k \leqslant N_\mathrm{ion}$.
For a given value \(K=k\), the number \(n\) of photons detected during a fluorescence-detection interval \(T_{\mathrm{det}}\) is assumed to follow a Poisson distribution:
\begin{equation}\label{2026_8_1_2}
\mathrm{Pr}(n\mid K= k)
=
e^{-\mu_k}\frac{\mu_k^n}{n!},
\qquad
n=0,1,2,\ldots,
\end{equation}
where \(\mu_k\) is the scheme-dependent mean photon count.

We impose two statistical requirements. First, for a scan containing $M=10^6$ frequency points, the expected number of false-positive candidate points must be smaller than five, namely, $M\alpha<5$, where $\alpha$ is the false-positive probability. Second, the false-negative probability at the adopted critical isomeric population must satisfy $\beta(P_{\mathrm{cri}})<0.01$. The effective total background rate is assumed to be $B=425~\mathrm{s}^{-1}$, comprising contributions from scattered light and detector dark counts. 

For the \(690\)- and 984-nm scheme, all ions initially contribute to the fluorescence signal. Nuclear excitation removes ions from the original fluorescence cycle. If \(k\) ions are transferred to the nuclear isomeric state, the remaining fluorescence rate is proportional to \(N_\mathrm{ion}-k\). The conditional mean photon count is therefore 
\begin{equation}\label{2026_8_1_3}
\mu_k=\left[B+R_{\mathrm{det}}\left(1-\frac{k}{N_\mathrm{ion}}\right)\right]T_{\mathrm{det}}.
\end{equation}

A positive nuclear-excitation event is identified when the detected photon count $N_{\mathrm{ph}}$ is at or below the threshold $N_{\mathrm{th}}$, namely, $N_{\mathrm{ph}}\le N_{\mathrm{th}}.$ In the absence of nuclear excitation ($k=0$), the mean photon count is $\mu_0=\left(B+R_{\mathrm{det}}\right)T_{\mathrm{det}}.$ The false-positive probability is therefore 
\begin{equation}\label{2026_8_1_4}
\alpha=\sum_{n=0}^{N_{\mathrm{th}}}\mathrm{Pr}\!\left(n\mid K=0\right)\equiv \mathrm{Pr}\!\left(n\le N_{\mathrm{th}}\mid K=0\right).
\end{equation}
A false-negative event occurs when the detected photon count remains above the threshold despite nuclear excitation. Its probability is
\begin{equation}\label{2026_8_1_5}
\beta(P)=\sum_{k=0}^{N_\mathrm{ion}}\mathrm{Pr}(K=k)\mathrm{Pr}\!\left(n>N_{\mathrm{th}}\mid K=k\right),
\end{equation}
where $\mathrm{Pr}\!\left(n>N_{\mathrm{th}}\mid K=k\right)\equiv 1-\mathrm{Pr}\!\left(n\le N_{\mathrm{th}}\mid K=k\right)$.

Using the detected fluorescence rates $R_{\mathrm{det}}$ evaluated at detunings of $5~\mathrm{MHz}$ for the 690-nm laser and $10~\mathrm{MHz}$ for the 984-nm laser (see Sec.~\ref{sec_2026_7_13_2}), together with the detection times specified in Sec.~\ref{SecTime}, we determine the photon-count thresholds and critical isomeric populations from Eqs.~\eqref{2026_8_1_4} and \eqref{2026_8_1_5} by imposing the two statistical requirements described above. The largest photon-count thresholds satisfying the false-positive constraint are found to be $N_{\mathrm{th}}=264$ and 263 for the 690- and 984-nm detection methods, respectively. The corresponding critical isomeric populations are about $P_{\mathrm{cri}}=45\%$ and $44\%$, respectively.

For the 1088-nm scheme, nuclear excitation increases the fluorescence signal. If \(k\) ions are in the nuclear isomeric state, the mean photon count is
\begin{equation}\label{2026_8_1_6}
	\mu_k	=	\left(	B+	R_{\mathrm{det}}\frac{k}{N_\mathrm{ion}}	\right)	T_{\mathrm{det}}.
\end{equation}
A positive nuclear-excitation event is identified when the detected photon count is at or above the threshold, namely, $N_{\mathrm{ph}}\ge N_{\mathrm{th}}$. In the absence of nuclear excitation, only the effective background contributes, and the mean photon count is $\mu_0=BT_{\mathrm{det}}$. The false-positive probability is therefore
\begin{equation}\label{2026_8_1_7}
\alpha=\mathrm{Pr}\!\left(n\ge N_{\mathrm{th}}\mid K=0\right).
\end{equation}
A false-negative event occurs when the photon count remains below the threshold despite nuclear excitation. Its probability is therefore
\begin{equation}\label{2026_8_1_8}
\beta(P)=\sum_{k=0}^{N_\mathrm{ion}}\mathrm{Pr}(K=k)\mathrm{Pr}\!\left(n< N_{\mathrm{th}}\mid K=k\right).
\end{equation}

Using the detected fluorescence rate $R_{\mathrm{det}}$ evaluated at a detuning of 30 MHz for the 1088-nm laser (see Sec.~\ref{sec_2026_7_15_1}), together with the detection time specified in Sec.~\ref{SecTime}, we determine the photon-count threshold and critical isomeric population from Eqs.~\eqref{2026_8_1_7} and \eqref{2026_8_1_8} by imposing the two statistical requirements. The smallest photon-count threshold satisfying the false-positive constraint is found to be $N_{\mathrm{th}}=32$. The corresponding critical isomeric population is approximately $P_{\mathrm{cri}}=42\%$.


%

\end{document}